\pdfoutput=1
\documentclass[namedreferences]{SolarPhysics}
\usepackage[optionalrh]{spr-sola-addons} % For Solar Physics 
\usepackage{graphicx}                    % For eps figures, newer & more powerfull
\usepackage{color}                       % For color text: \color command
\usepackage{url}                         % For breaking URLs easily trough lines
\usepackage{lscape}
\usepackage{rotating}
\begin{document}

\begin{article}

\begin{opening}

\title{Origins of Rolling, Twisting and Non-Radial Propagation of Eruptive Solar Events}

\author{Olga~\surname{Panasenco}$^{1}$ \sep Sara F.~\surname{Martin}$^{1}$
\sep Marco~\surname{Velli}$^{2}$ \sep Angelos~\surname{Vourlidas}$^3$ }

%%%%%%%%%%%%%%%%%%%%%%%%%%%%%%%%%%%%%%%%%%%%%%%%%%%
%% Runningheads
%

\runningauthor{Panasenco, Martin, Velli and Vourlidas}
\runningtitle{Roll Effect and Non-Radial Propagation}

%%%%%%%%%%%%%%%%%%%%%%%%%%%%%%%%%%%%%%%%%%%%%%%%%%%
%% Affiliations 

\institute{$^1$ Helio Research, La Crescenta, CA, 91214, USA \\ 
email: \url{panasenco.olga@gmail.com} \\
$^2$ Jet Propulsion Laboratory, California Institute of Technology, Pasadena, CA, USA\\
$^3$ Space Sciences Division, Naval Research Laboratory,
Washington DC, USA}
%%%%%%%%%%%%%%%%%%%%%%%%%%%%%%%%%%%%%%%%%%%%%%%%%%%
%%% Abstract 

\begin{abstract}
We demonstrate that major asymmetries in erupting filaments and CMEs, namely
major twists and non-radial motions are typically related to the larger-scale ambient environment around eruptive events. Our analysis of prominence eruptions observed by the STEREO, SDO and SOHO spacecraft shows that prominence spines retain, during the initial phases, the thin ribbon-like topology they had prior to the eruption. This topology allows bending, rolling, and twisting during the early phase of the eruption, but not before.
The combined ascent and initial bending of the filament ribbon is non-radial in the same general direction as for the enveloping CME. However, the non-radial motion of the filament is greater than that of the CME. In considering the global magnetic environment around CMEs, as approximated by the Potential Field Source Surface (PFSS) model, we find that the non-radial propagation
of both erupting filaments and associated CMEs is correlated with the presence of nearby coronal holes, which deflect the erupting plasma and embedded fields. In addition, CME and filament motions respectively are guided towards weaker field regions, namely null points existing at different heights in the overlying configuration. Due to the presence of the coronal hole, the large-scale forces acting on the CME may be asymmetric. We find that the CME propagates usually non-radially in the direction of least resistance, which is always away from the coronal hole. 
We demonstrate these results using both low and high latitude examples. 
\end{abstract}


\end{opening}

\section{Introduction}

Eruptive solar events which include a coronal mass ejection (CME), an erupting filament with surrounding cavity and a flare are distinguished from \emph{simple confined} flares which lack most of these erupting signatures. 
The term \emph{CME} is increasingly used in reference to the whole eruptive sequence and not just the ejection of coronal mass.
Regardless of definition, there has been widespread recognition, for over three decades, of the close association among CMEs, erupting filaments, coronal cavities and flares.

What do CMEs and erupting filaments have in common? All of these events originate above and around regions where the magnetic field changes polarity, seen in photospheric magnetograms and variously named a neutral line or polarity inversion line or a polarity reversal boundary. We know that erupting filaments, in particular, originate from specific polarity reversal boundaries that are also \emph{inside filament channels} (Martin,
1990; Gaizauskas, 1998). We also already know they only come from filament channels that have reached maximum development; i.e. have reached maximum magnetic shear along the polarity boundary (Martin, 1998). Therefore, not every polarity reversal boundary is a \emph{filament channel}. 

A filament channel is a volume of space around (and encompassing) a polarity reversal boundary with maximum magnetic shear.  A filament is not necessarily present but the channel is a necessary condition for the formation of a filament.
The formation of filament channels in the chromosphere
has been described by Smith (1968), Foukal (1971), Martin (1990), Gaizauskas et al. (1997), Wang and Muglach (2007) and Martin, Lin, and Engvold (2008) and Martin et al. (2008). In and around active regions or decaying active regions observed in H$\alpha$ a filament channel may be recognized by the  presence of fibrils aligned along a polarity reversal boundary and the local magnetic field.  Because no fibrils cross the polarity reversal boundary in a fully-developed filament channel and fibrils are field-aligned, one may conclude that no magnetic field lines from
active region or network magnetic fields cross this  boundary in the chromosphere, above or within filaments. This is also true for the low corona. The initial lack of coronal connectivity across polarity reversal boundary inside the filament channel volume, under, across and along the filament is confirmed by  Wood and Martens (2003). We find that the coronal cells observed and modeled by Sheeley and Warren (2012), do not cross the polarity reversal boundary within a filament channel at the heights below the filament spine top (Figure 1). Coronal cells originate from the network field concentrations and show the same patterns as chromospheric fibrils because they follow the same filament channel magnetic topology, with a (presumably) strong horizontal component of the field.

\begin{figure}
\center
\includegraphics[scale=.21]{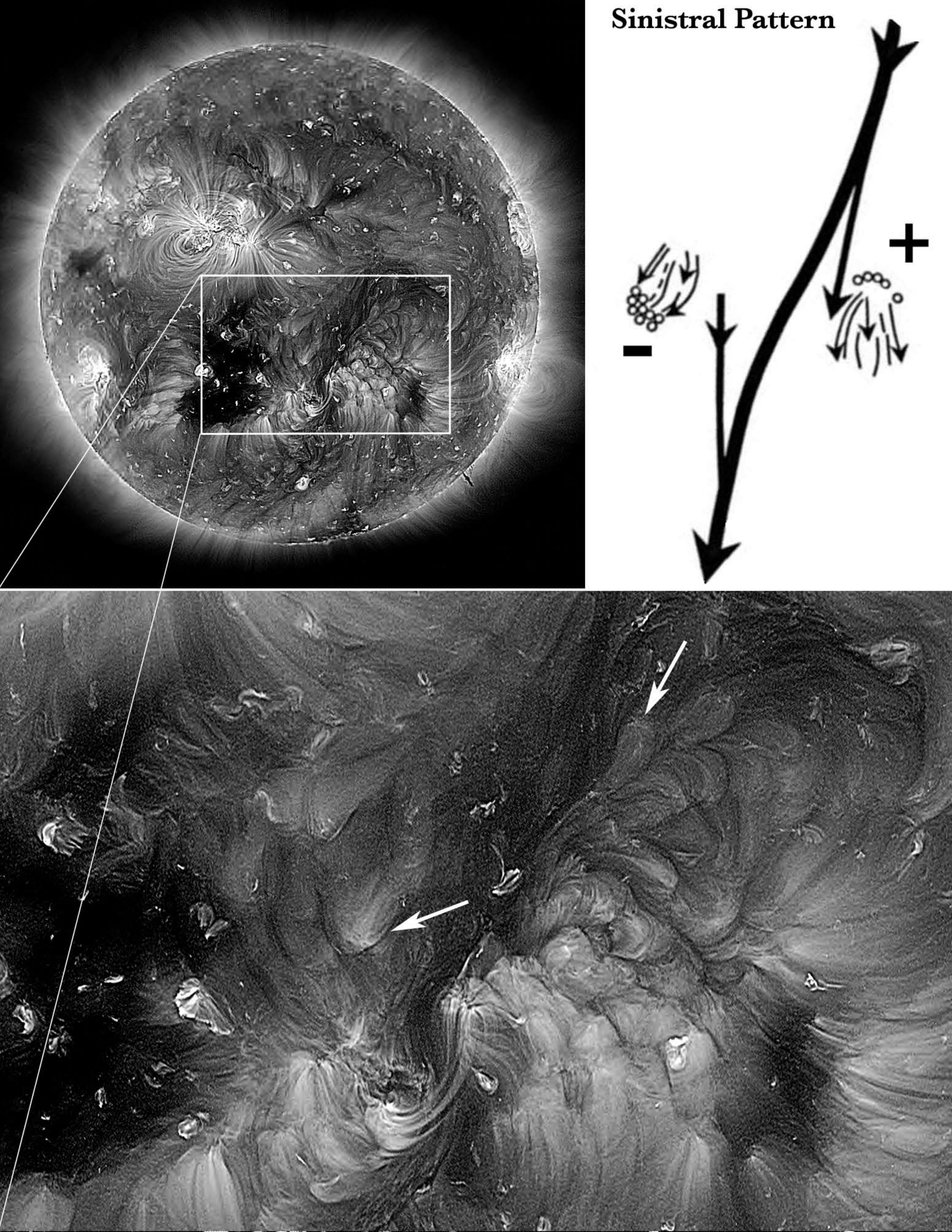}
\caption{Shown here are coronal cells on both sides of the filament channel (two representative cells are indicated by the white arrows), note that the cells on opposite sides of the channel have have cusp-shaped tops that are in opposite directions: observe that they do not cross the channel but follow the same pattern as chromospheric fibrils within a sinistral filament channel (a left-handed filament channel) as depicted in the  schematic representation  in the upper right where a sinistral filament and adjacent chromospheric fibrils with a sinistral pattern are shown (Martin, Lin, and Engvold, 2008).  Coronal cells form at heights $\sim$ 6-10 Mm (Sheeley and Warren, 2012), much lower than average filament heights $\sim$ 50 -70 Mm. The SDO/AIA 193 {\AA} image is from 19 June 2011 05:39 UT. } 
\end{figure}

As deduced from chromospheric fibrils and coronal cells, the direction of the network magnetic fields associated with chromospheric fibrils and coronal cells changes from upward on
one side of a polarity reversal boundary to horizontal along the polarity reversal boundary and to vertically downward on
the other side of the polarity boundary.  A vital precondition for the formation of a filament channel 
is a magnetic field with a strong horizontal component (Gaizauskas, 1998) and the chromospheric fibrils 
 in a filament channel trace such a horizontal pattern along its polarity reversal boundary. So far no observational evidence of the existence of an ascending twisted flux rope at the chromospheric level within a filament channel
above a polarity reversal boundary has been found; this would require fibrils to systematically change their orientation and cross the polarity reversal boundary. Conversely, were the reverse to be true, emerging, twisted flux tubes playing a major role in the physics of filaments and filament channels, one would expect there to be ubiquitous evidence of polarity crossing fibrils, which would necessarily appear earlier, if there was a flux rope axis submerged in the photosphere (Wood and Martens, 2003).
 
This is a fundamental observation which allows one  to deduce at least one significant moment in the formation of  CMEs: the most likely time of initiation of twist in CMEs. There is evidence that once an eruption is underway, a twisted flux rope does indeed form during magnetic reconnection of the stretched, sheared overlying coronal loop system and concurrently with the initiation of flare ribbons in the chromosphere (Wang, Muglach and Kliem, 2009).
In other words, the observational evidence suggests that the formation of the flux rope of CMEs occurs after the filament and  coronal loop system are already ascending. 

Similar to the overlying coronal loops within a coronal loop system, before eruption, filaments rise slowly.  Unlike coronal loops and if  a filament erupts non-radially, as frequently happens, the top of its spine
first bends to one side and evolves into a sideways rolling motion, known as the roll effect. As shown by 304\AA\ observations
from SOHO and STEREO and earlier H$\alpha$ Doppler observations, the rolling motion unambiguously propagates down
the legs of erupting filaments resulting in the large scale twists which are commonly observed in them. The rolling of the spine initiates structural twist with opposite chirality in the two legs (Figure 2). Concurrently, rotational mass motions with the opposite signs of chirality (and helicity) in the two legs are observed to flow downward while the body of the filament ascends.  However, in asymmetric eruptions, both legs are not always observed (Martin, 2003; Panasenco and Martin, 2008).

\begin{figure}
\center
\includegraphics[scale=0.24]{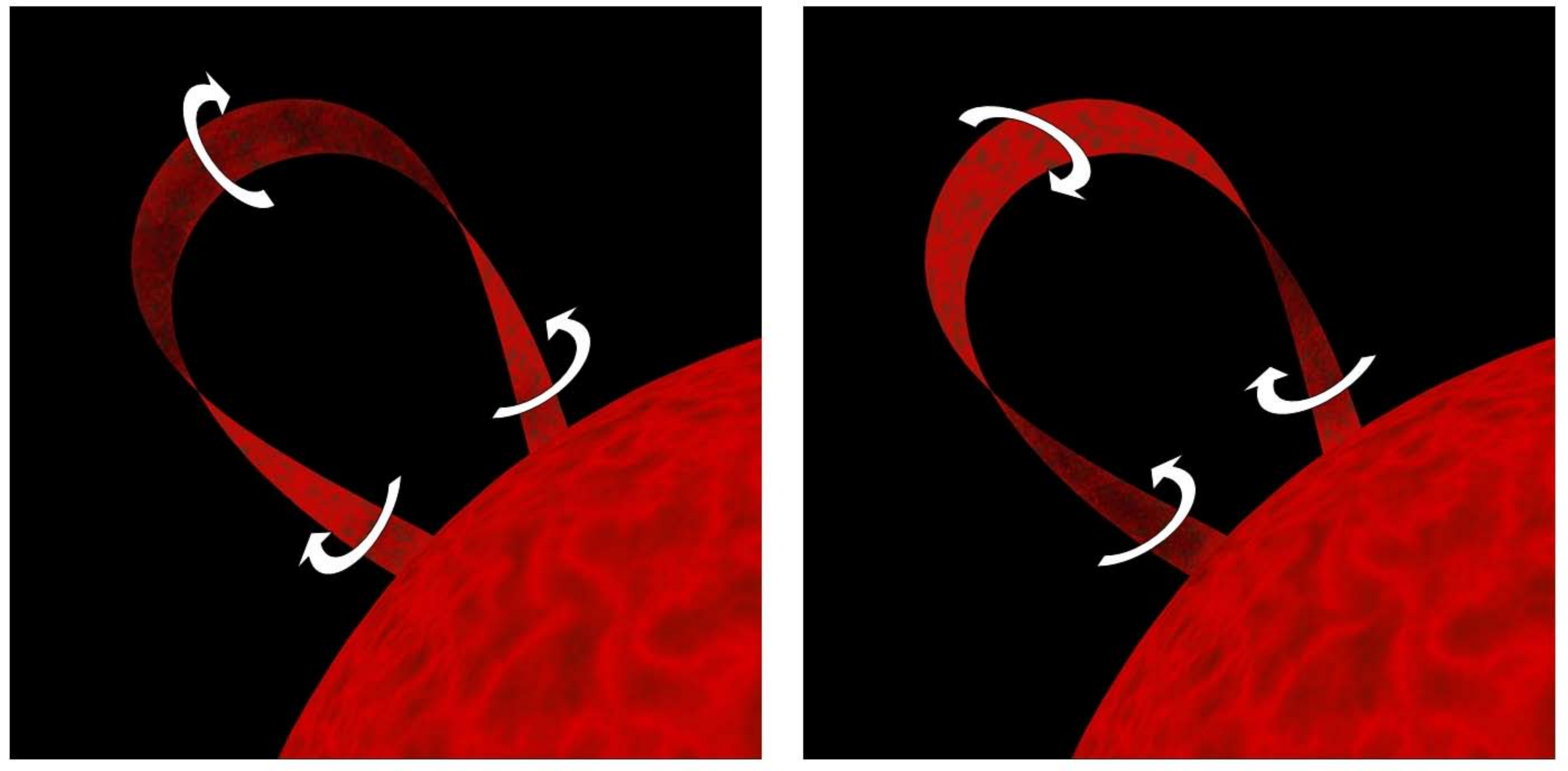}
\caption{ Schematic of the two forms of the roll effect in erupting prominences. The roll effect initiates at the top of a filament and propagates downward into the legs. The legs respond by twisting. Rotational motions of the legs depend on the roll direction. There is no photospheric motion fast enough to drive the observed rotation of the legs.}
\end{figure}

In addition to the observed absence of twist in the pre-eruptive state, further evidence that the energy creating the twist comes from above was found in
Doppler shifts (Martin 2003); the rotational motions in the legs of erupting filaments are not only opposite in sign to each
other but the twists in both legs are opposite in sign to that required if the observed sense of twist were
generated at the feet or in the legs of the erupting filament (Martin 2003, Panasenco and Martin, 2008). 

Here we will examine events showing that the combined ascent
and initial bending of the filament is non-radial in the same general direction as for the surrounding CME. 
Also, the non-radial motion of the filament is \emph{greater} than that of the CME. By considering the global magnetic environment around CMEs, as can be approximated by the Potential Field Source Surface (PFSS) model, we will see 
that both erupting filaments and their surrounding CMEs are non-radial moving away from nearby coronal holes and toward null points. Due to the presence of the coronal hole, both
the local forces on an erupting filament and the global forces on the CME are asymmetric. That the CME propagates non-radially always away from coronal holes was demonstrated in many studies (e.g., Plunkett et al., 2001; Gopalswamy et al., 2003; Cremades and Bothmer, 2004; Cremades, Bothmer and Tripathi, 2006; Gopalswamy et al. 2009; Kilpua et al., 2009; Zuccarello et al., 2012). Prominence deflection and rolling motion during eruptions received less attention, though there has been some work in the same general direction as advocated here (Filippov, Gopalswamy and Lozhechkin, 2001;  Martin 2003; Panasenco and Martin, 2008; Bemporad 2009; Panasenco et al., 2011; Pevtsov, Panasenco and Martin, 2012; Liewer, Panasenco and Hall, 2012). The formation of the CME flux rope appears to occur in the early phase of filament eruption, when the rolling motion of the filament is already in progress (Liewer, Panasenco and Hall, 2012).

The roll effect appears very early during the beginning phase of prominence eruptions. We will show that it is caused by an asymmetry  in magnetic structure and magnetic flux density on the two sides of the filament. The non-radial direction of the eruption of the whole filament system is caused directly either by the open coronal hole magnetic field near the filament channel (cases studied in this paper), or by other strong magnetic field configurations which might be in the neighborhood of the eruption (strong in reference to the overall energy of the eruption).

The paper is organized as follows: we begin with a general introduction to the coronal structures which lead to eruptive solar events.  We then discuss a series of recent eruptions within such structures, which may be referred to generically as either streamers or pseudostreamers, exhibiting non-radial expansion and propagation.  We then correlate such behavior with the properties of the ambient corona. We also discuss the three-dimensional structure of such events, focusing on one particular case where the two well separated STEREO viewpoints show that what appears to be a highly kinked, flux rope-like erupting prominence and cavity is a projection effect.

\section{Magnetic topologies overlying filament channels}

Filament channels and filaments develop along and above polarity reversal boundaries in the photosphere  and lie under coronal loop systems also known as coronal arcades. In the instances most commonly studied, these arcades correspond to  a helmet-streamer type structures in the outer corona. The outer, weaker fields do not always close over the arcade but may remain open, oppositely directed adjacent fields (what we will refer to as a globally dipolar structure). Recently, filament channels and filaments lying below arcades and belonging to more complex magnetic structures have been shown to be very important for multiple and global solar eruptive events, such as that of  01 August 2010  (Schrijver and Title, 2011; T{\"o}r{\"o}k et al., 2011). The prominences in that event had a magnetic structure corresponding to a hierarchy of higher order multipolar structures resulting in a pseudostreamer (Wang, Sheeley and Rich, 2007), which was found to contain \emph{twin} filaments at its base (Panasenco and Velli, 2010).  Such twin filaments are topologically connected, sharing a neutral point and a separatrix dome. This was a case in which two polarity reversal boundaries contain between them fields with a polarity opposite to that of the global unipolar configuration surrounding them (\emph{tripolar} pseudostreamer). 

The magnetic field, based on the potential field source surface (PFSS) reconstruction, shows that field lines of the arcades above filaments are often asymmetrically placed with respect to the polarity reversal boundaries, because the overarching magnetic field is stronger at one side. In the PFSS this is shown by the greater density of field lines on one side of the arcade with respect to the other. As we shall see, this is a tell-tale sign for strong rolling and nonradiality of the eruption.  The question naturally arises as to the fidelity of the PFSS model in the neighborhood of filaments and filament channels, where the strong shears naturally lead to the formations of currents in the corona, currents which are also required to get to the instability involved in the filament eruption and CME formation. In reality, the PFSS does a good job because at least a component of the axial field along the filament spine is potential in nature. This may be understood as having to do with the finite longitudinal extent of the channel Ð so that the polarities on opposite side of the Êinversion line are not translationally invariant along it. Indeed, this component is recovered also by photospheric shear and flux transport models which correctly describe filament channel formation statistically based on photospheric transport (Yeates, Mackay and van Ballegooijen, 2007). In the following sections we will make use of the PFSS model of Schrijver and DeRosa (2003)  to study the coronal magnetic field and its role in determining the trajectory of filaments and CMEs for a whole series of events including even higher order multipoles below streamers and pseudostreamers.
 
 \begin{figure}
\center
\includegraphics[scale=0.12]{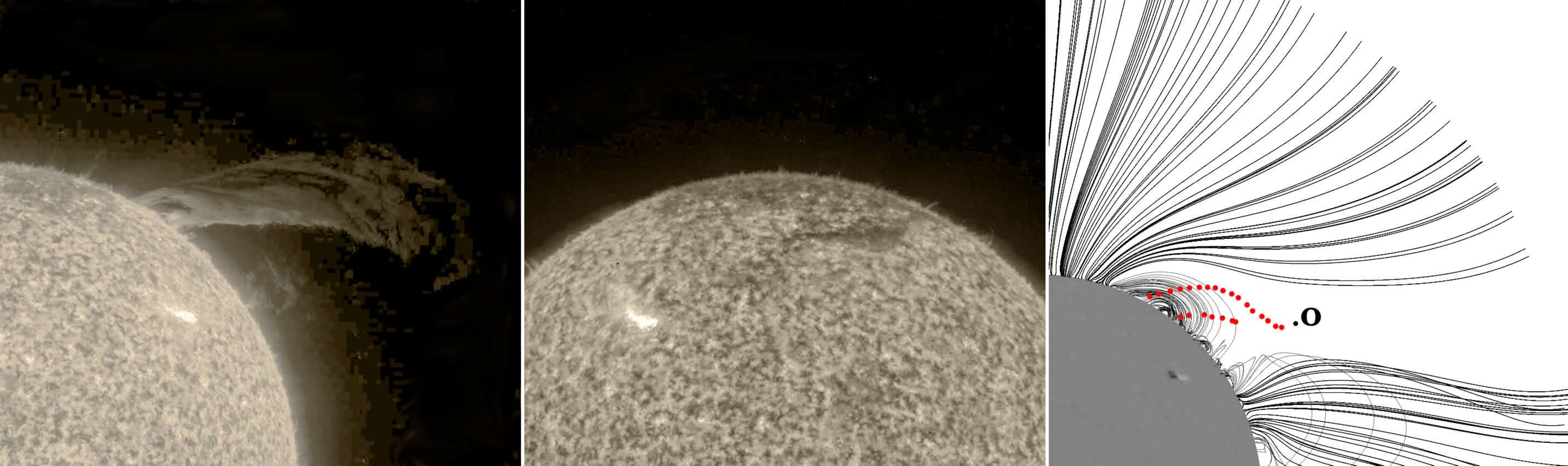}
\caption{ Non-radial eruption with the roll on 2 November 2008.  The large curvature results from sideways rolling of the top of the prominence. Left and middle are STEREO-B and STEREO-A/EUVI  304 \AA\  images of the prominence during the eruption at 01:57:04 and 01:56:15 UT respectively. Right image is the PFSS extrapolation of the open magnetic field lines (pseudostreamer) and the coronal loops overlying the filament before the eruption on 1 November 2008 12:04 UT (rotated to the limb view); red dotted lines represent the erupting filament directed toward the null point, O.}
\end{figure}

\begin{figure}
\center
\includegraphics[scale=0.115]{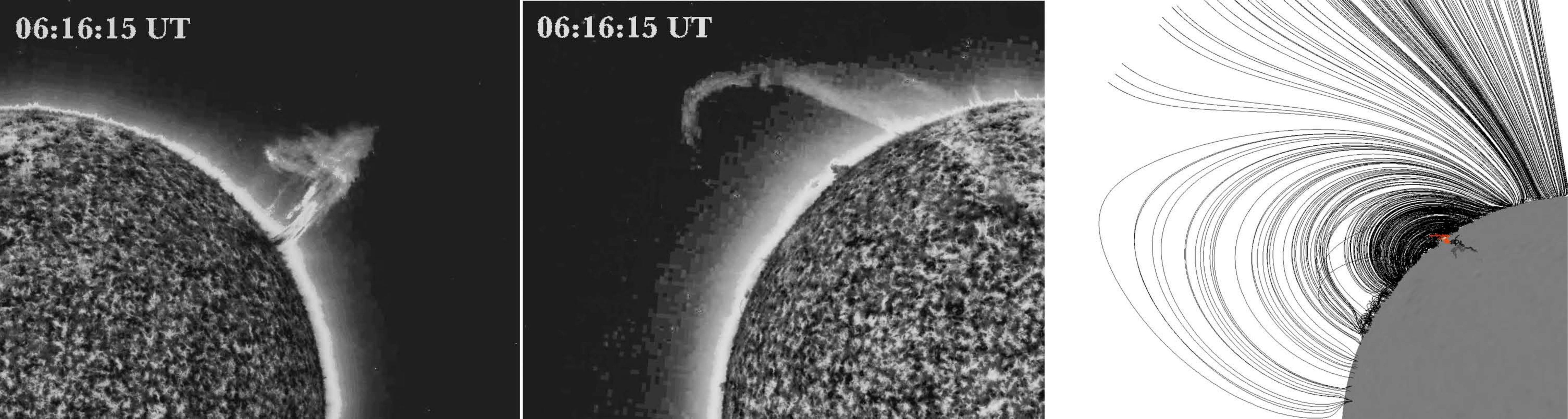}
\caption{ Non-radial eruption on 12 December 2008.   Left and middle are STEREO-B and STEREO-A/EUVI  304 \AA\  images of the prominence during the eruption. Right image is PFSS extrapolation of the open solar magnetic field and the coronal loops overlying the filament (STEREO-A/EUVI  304 \AA\ ) before its eruption. The  position of the filament under the arcade is very asymmetric.}
\end{figure}

\begin{figure}
\center
\includegraphics[scale=.2]{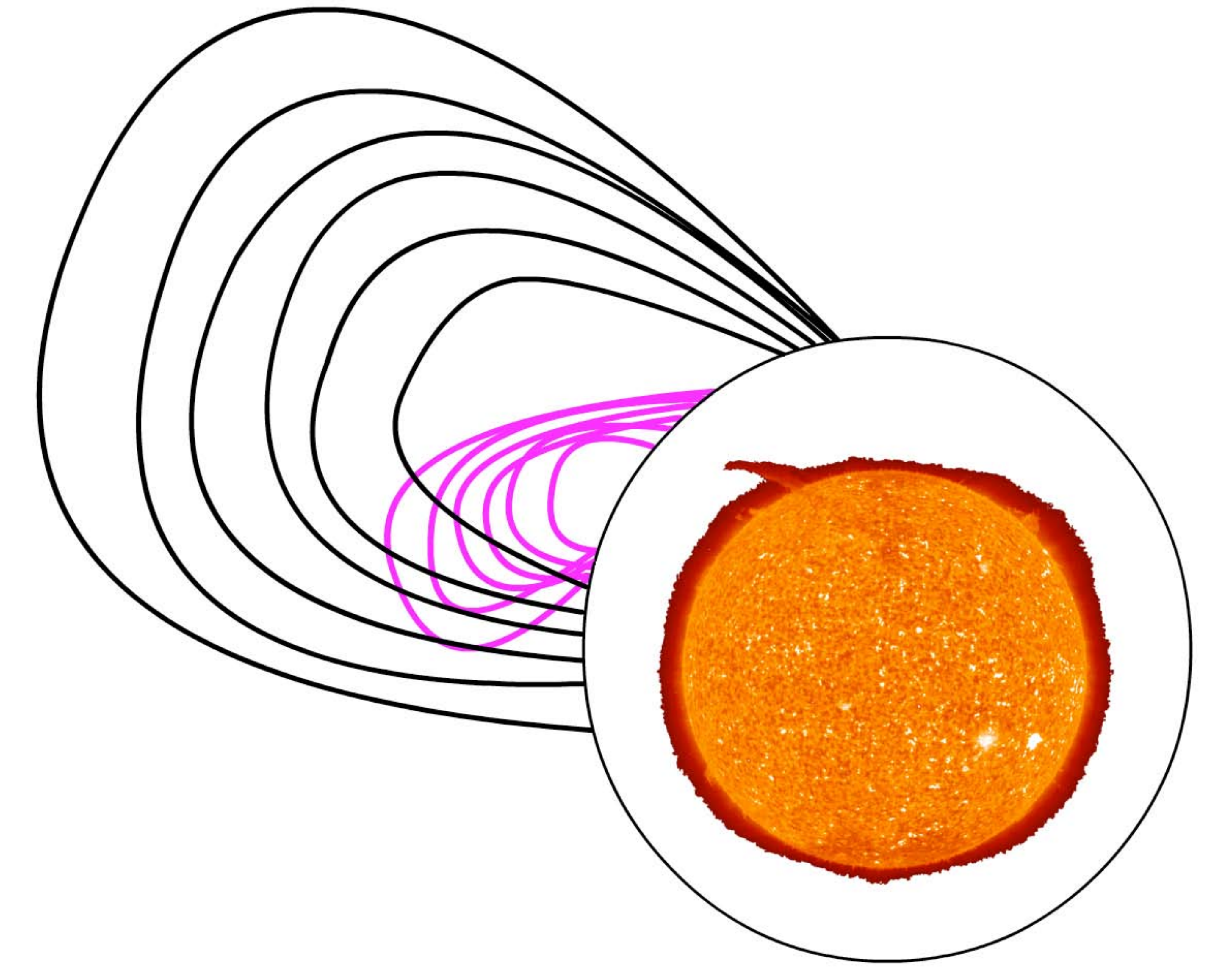}
\caption{ Non-radial eruption on 12 December 2008.  Superposition of the outer boundaries of CME (black lines)  and erupting prominence inside (magenta lines) as observed by STEREO-A/COR1 during different moments of the eruption and  the solar disk  with the prominence before eruption (STEREO-A/EUVI  304 \AA\ ). }
\end{figure}

\section{Non-radial eruptions with the roll effect}
We consider here a set of prominence and CME eruptions which display both the roll effect and significant non-radiality. The different examples paint
a picture of the rich variability in motions which the prominence and CME may display, depending on the detailed magnetic field configuration surrounding the eruption and the global coronal context.  We use images from the \emph{Sun-Earth Connection Coronal and Heliospheric Investigation} (SECCHI: Howard et al., 2008) aboard the \emph{Solar-Terrestrial Relations Observatory} (STEREO: Kaiser et al., 2008) and from the \emph{Atmospheric Imaging Assembly} (AIA: Lemen et al., 2012) aboard the \emph{Solar dynamics Observatory} (SDO: Pesnell et al., 2012). We also use information from the \emph{Large Angle Spectrometric Coronagraph} (LASCO; Brueckner et al 1995) on the \emph{Solar and Heliospheric Observatory} (SOHO: Domingo et al., 1995).

\subsection{2 November 2008 filament eruption }
This strongly non-radial eruption (Figure 3) was described in detail by Kilpua et al. (2009). The right panel in Figure 3 shows that the filament channel was located under the northernmost arcade of a double arcade system underneath a pseudostreamer, a large-scale coronal structure which differs from the  classical helmet streamer because the magnetic field is unipolar above the dome like closed field regions. The pseudostreamer itself, adjacent to a strong northern coronal hole, expanded outward with considerable equatorward deflection. At the early stage of the eruption, the motion of the filament was directed towards the local null point, O, below the pseudostreamer separatrix (Figure 3). 

\subsection{12 December 2008 filament eruption }
This event, illustrated in Figure 4,  is described in detail in the paper by Panasenco et al., 2011. It was observed by both STEREO-A, at the north east limb (middle panel), and STEREO-B, at the north west limb (left panel). Though the spacecraft were separated at the time by  $86.7^{\circ}$, both imagers show the rolling motion of the erupting prominence. The erupting prominence and CME were deviated from a radial trajectories by different degrees. To illustrate this, in Figure 5 we superimpose outlines  of the outer boundaries of the CME and the bright core of the prominence plasma for different consecutive moments during the eruption (as observed by the \emph{Extreme Ultraviolet Imager} (EUVI-A)  304 \AA\  and COR1-A, respectively). The deviation of the CME from radial propagation is about   $40^{\circ}$, for the prominence Ð about $60^{\circ}$ and the difference between the prominence and CME central direction of the eruption is about $20^{\circ}$  for this STEREO-A view. Since the prominence erupts approximately toward STEREO-B, and the separation angle between A and B is close to  $90^{\circ}$, we may safely assume that the projection on the sky in the STEREO-A view shows a separation of trajectories which is smaller than the real separation between the propagation directions for  prominence and CME, shown in Figure 5.

\begin{figure}
\center
\includegraphics[scale=0.13]{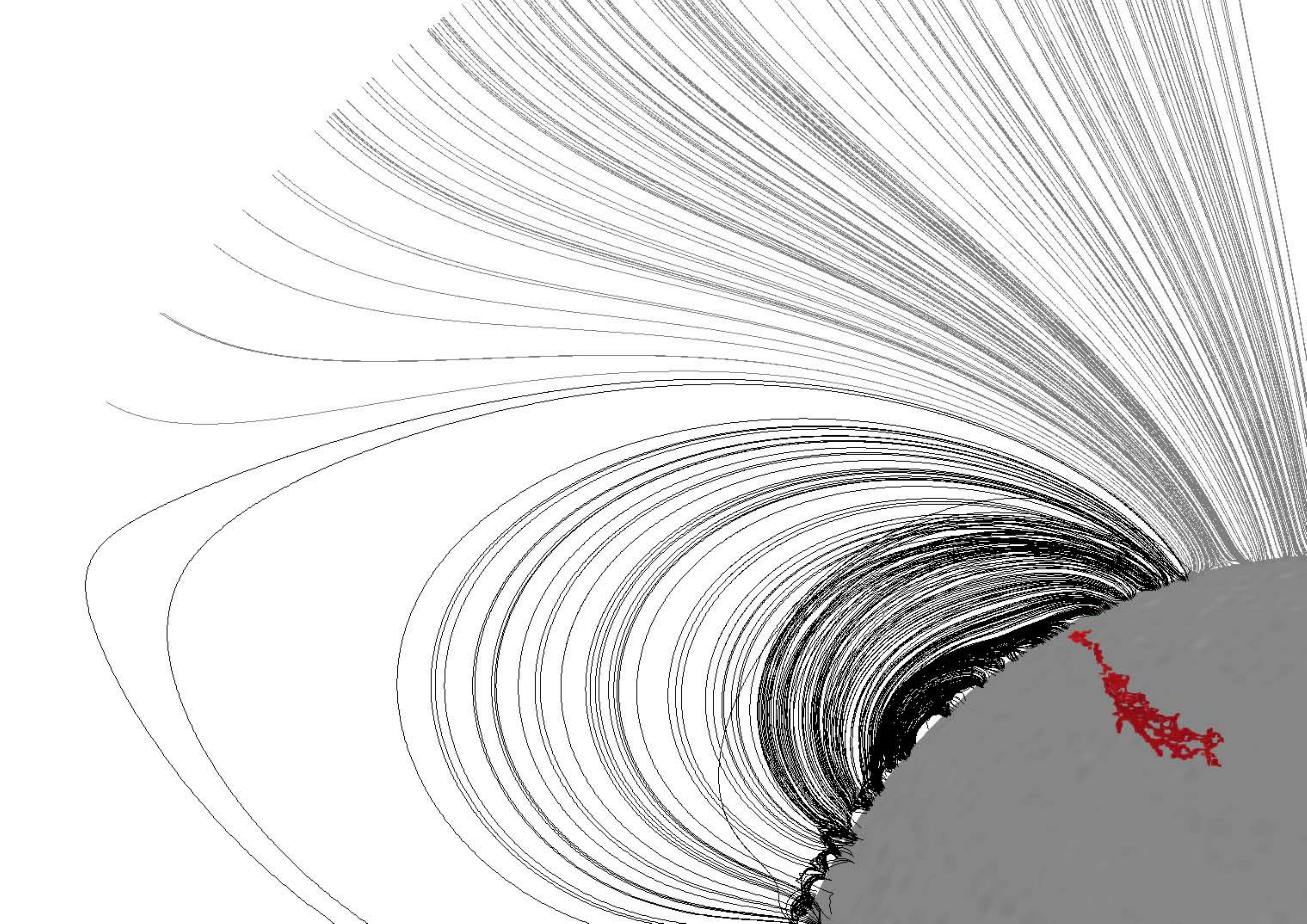}
\caption{ Magnetic configuration above the filament  before its eruption on  26 September 2009 (STEREO-B).  The asymmetry in the position of the filament under the arcade is very strong.}
\end{figure}

\begin{figure}
\center
\includegraphics[scale=0.3]{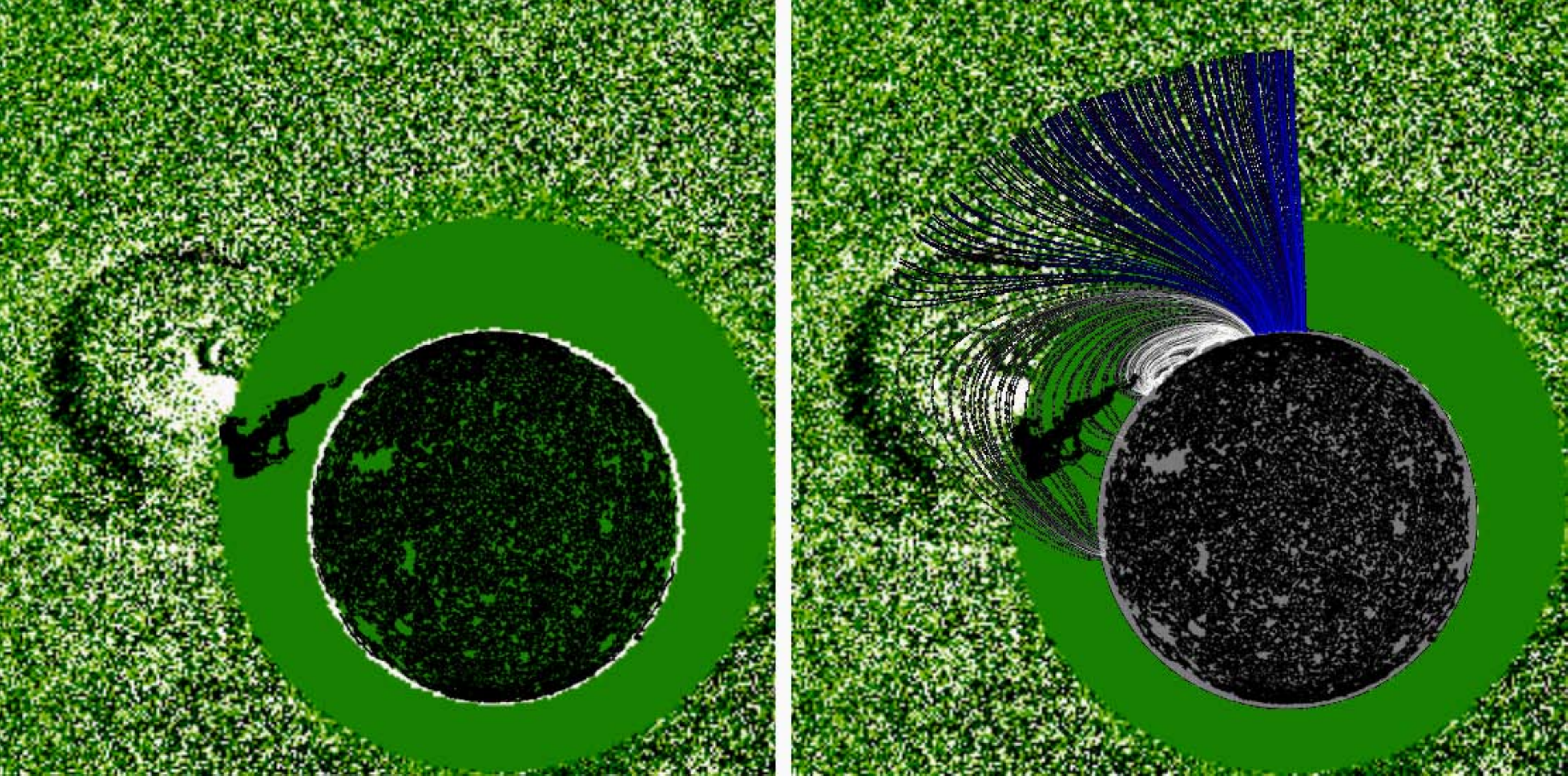}
\caption{ Non-radial eruption on 26 September 2009 at  22:45 UT.  At the left: a superposition of the COR1-A and EUVI-A 304\AA\ images; at the right the same with added PFSS reconstruction. The non-radial motion of the filament is much greater than that of the CME.}
\end{figure}

\subsection{26 September 2009 filament eruption}  
This eruption was observed by both STEREO, at the north east limb by A, and across the disk by B as shown in Figure 7 . The separation angle between the spacecraft was  $117.4^{\circ}$. EUVI-B 304 \AA\ images show the prominence crossing the disk during the eruption. STEREO-A observed this eruption above the limb. The position of the filament relative to the coronal arcade and open field of the coronal hole before the eruption is shown in Figure 6. The asymmetric location of the filament is very pronounced. Figure 7 shows the eruption on 26 September 2009 at 22:45 UT as observed simultaneously by STEREO-A/COR1 and EUVI instruments. The non-radial motion of the erupting filament is much greater than that of the CME.  The CME envelope expands, containing the prominence within it, but the two bodies are not rigidly linked to one another. Plasma of the prominence moves approximately along the line which can be described as a tangent line to the limb from where the prominence erupted. Apparently the imbalance in magnetic pressure under the coronal arcade and around the filament  during the early stage of the eruption was such that the stronger field on the coronal hole side of the arcade caused such an extensive lateral motion of the filament. The corresponding CME also deviates from radial propagation. Analyzing the projection on the sky as viewed by STEREO-A we found that the difference between the prominence and CME deviations amounted to about $30^{\circ}$. Figure 8 shows an intermediate projection of the eruption on the plane of the sky as observed by the SOHO/EIT and LASCO C2 instruments. This projection viewpoint is midway between those of STEREO-A and B, therefore the apparent difference in the deviation angle for the prominence and CME is smaller, about  $20^{\circ}$.

\begin{figure}
\center
\includegraphics[scale=0.2]{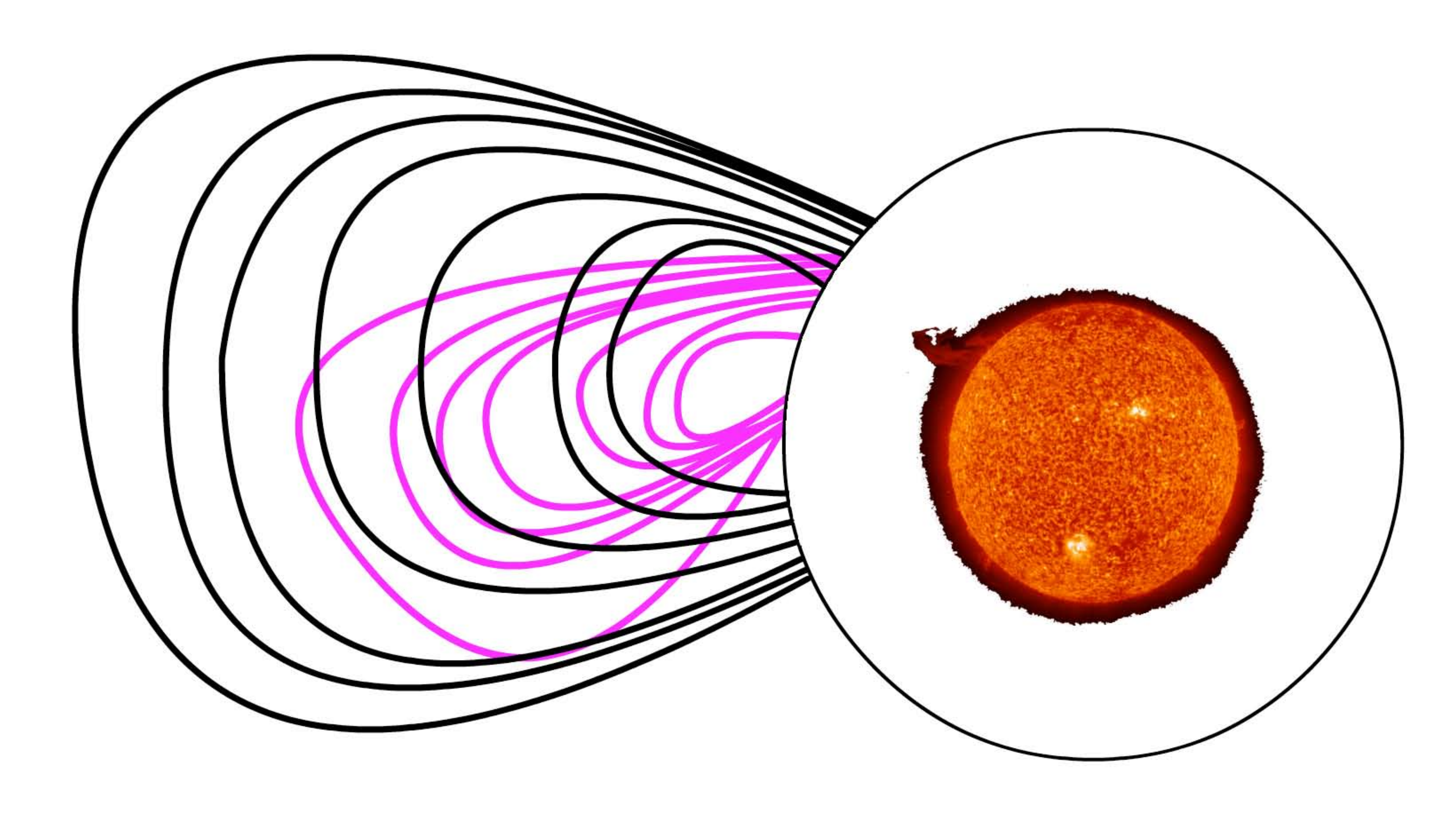}
\caption{ Non-radial eruption on 26 September 2009.  Superposition of the outer boundaries of the CME (black lines)  and erupting prominence inside (magenta lines) as observed by SOHO/LASCO C2 during different moments of the eruption (26-27 September) and  the solar disk  with the prominence before eruption (SOHO/EIT 304 \AA\ ).}
\end{figure}

\subsection{30 April 2010 filament eruption.}

\begin{figure}
\center
\includegraphics[scale=0.2]{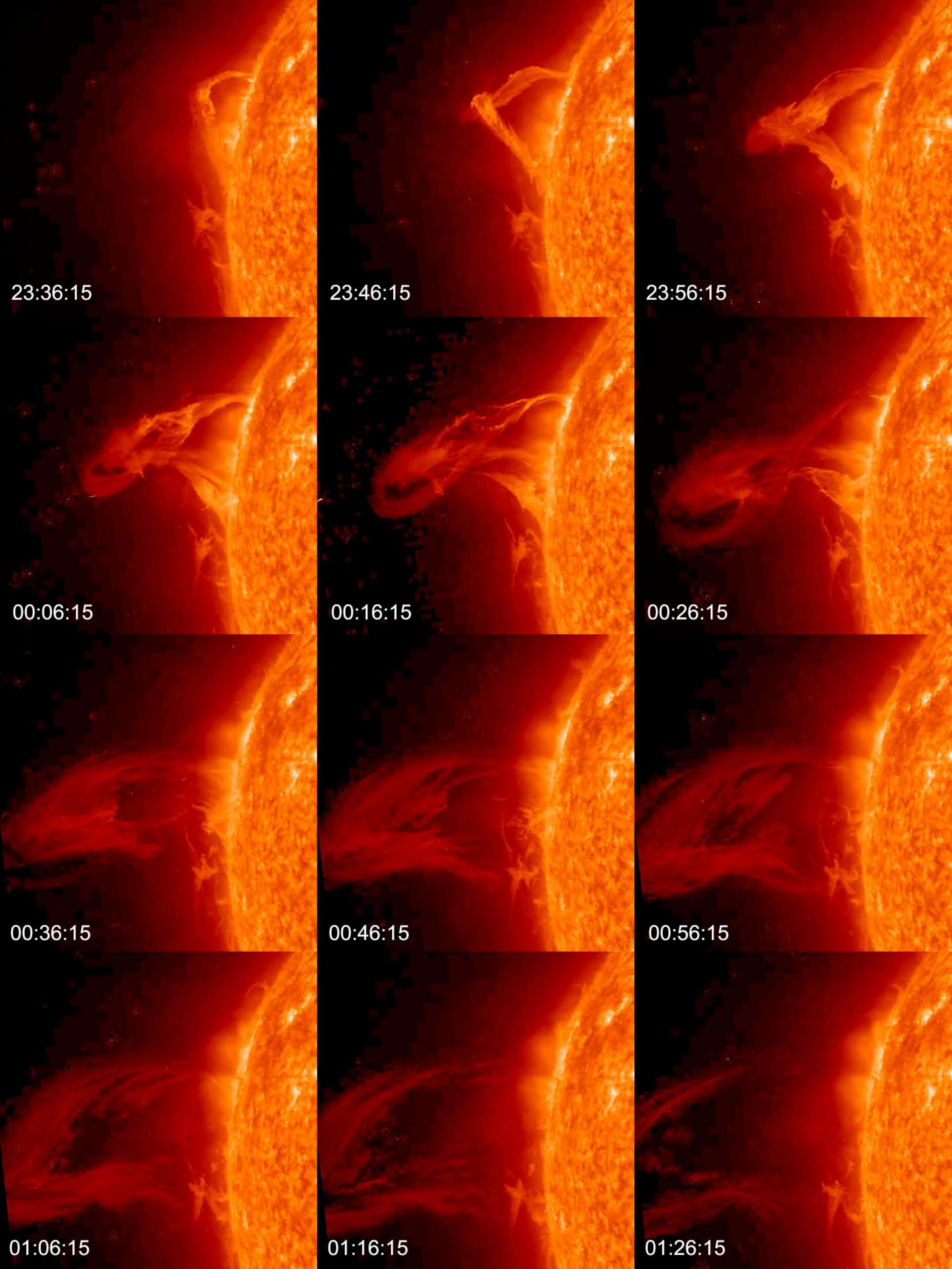}
\caption{ Non-radial eruption on 30 April - 1 May 2010 as observed by STEREO-A/EUVI 304 \AA\  (the simultaneous observation of the eruption from the ground-based observatory (MLSO) is in the supplementary material).}
\label{20100430}
\end{figure}

This non-radial filament eruption with the roll effect was observed against the solar disk by the Mauna Loa Solar Observatory (MLSO) on 30 April 2010. The movie of the first 40 min of the eruption (23:25 - 00:05 UT) in the He I spectral line shows that the filament crossed the disk and the neighboring active region in the south-eastern direction. This direction of the eruption was away from the coronal hole north-west of the filament channel (movie available in the supplementary material). The distance which the filament covered over these 40 min was about 160000 km, corresponding to a projected speed of $\sim$ 65 km s$^{-1}$. The first four frames in Figure 9 correspond to this period but observed from the different point of view of STEREO-A . The top of the prominence is bending  away and to the left of the viewing direction of STEREO-A (Figure 9) creating an apparent kink and twist in the prominence ribbon. However, the apparent kink crossing may be an illusion from viewing the warped ribbon of erupting filament mass seen in a 2-dimensional image. MLSO and STEREO-B observations show the filament spine as straight, without any twisting, crossing the disk. The apparent kink disappears as seen in the final six frames in Figure 9.  The images are suggestive of a kink structure until 00:26 UT. With the next 304 \AA\ image (at 00:36 UT) the apparent kink disappears as the eruption proceeds along the southeast extension of the filament channel (final six frames in Figure 9). 

T{\"o}r{\"o}k, Berger and Kliem (2010) have pointed out examples of kink unstable flux ropes where there is no crossing as seen from angles. While this may be the case, additional considerations may come from the properties of the line-tied kink instability. The kink instability requires, for a flux rope, a certain minimum number of turns that the magnetic field must make from end to end, not on the axis of the flux rope configuration, but in its immediate surroundings.  Oftentimes this is described in terms of the ratio of the average poloidal flux (i.e. the flux circulating around the axis of the flux rope) to the axial flux (the integrated field along the flux rope axis or spine). The required twist for kink instability is at least 2.5$\pi$ and depending on the configurations considerably more.  This is more than one complete turn of the field in the domain (Hood and Priest, 1979; Velli, Hood and Einaudi, 1990) and is a prerequisite,  independent of observed kinking of the spine. If the spine is bent, and there is no tracer of plasma making one complete turn around the spine, the presumed axis of the prominence,  the evidence for kinking becomes weak. The introduction of toroidal geometry (and the hoop force) does not change these constraints on the kink instability, i.e the flux ropes of T{\"o}r{\"o}k, Berger and Kliem (2010) also have field lines wrapping around the axis for more than one complete turn. Because individual field lines are the only real relevant concept once line tying is taken into account (because the concept of a flux surface is lost if lines end in different regions in the photosphere), rather than flux surfaces, plasma tracers along such turning field lines should never disappear no matter what the point of view.

The position of the filament under its coronal arcade before its eruption was very asymmetric. Figure 10 shows the PFSS extrapolation of the magnetic field lines above the filament on 2010 Apr 30 18:04 UT. This asymmetry is thought to lead to asymmetric forces on the filament which cause the rolling motion of the filament at the early stages of the eruption, and create a condition for the non-radial propagation of the corresponding CME (Figure 11).  The right image in Figure 11 shows the CME front and position of the prominence at the limb from STEREO-A; the CME non-radial propagation relatively to its point of origin is clearly visible.

\begin{figure}
\center
\includegraphics[scale=0.22]{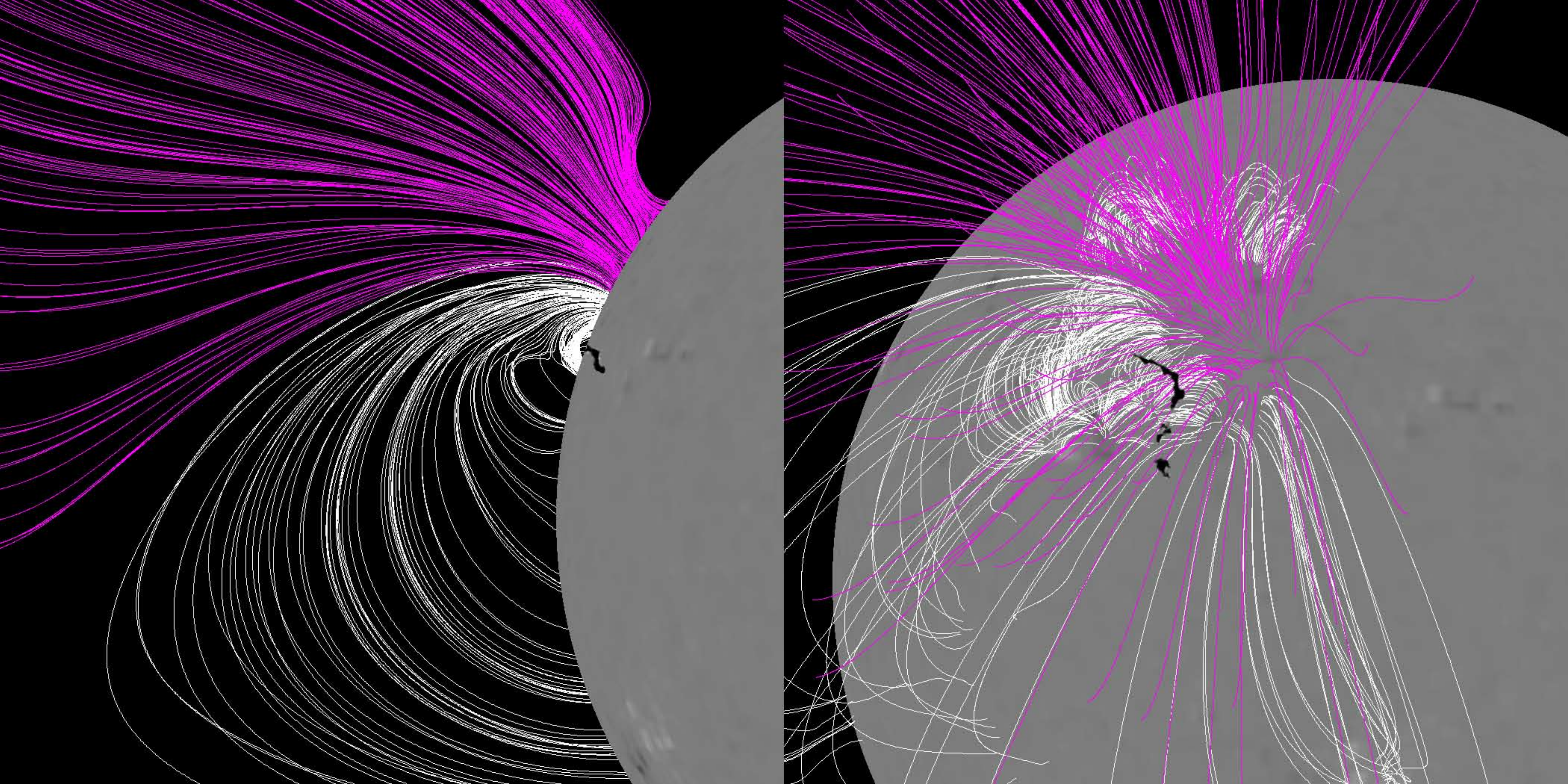}
\caption{ Superposition of the H$\alpha$ filament and the PFSS reconstruction of the filament arcade and the open field of the neighboring coronal hole on 30 April 2010 18:04 UT. Left: limb view; right: normal view. }
\label{20100430_pfss}
\end{figure}

\begin{figure}
\center
\includegraphics[scale=0.22]{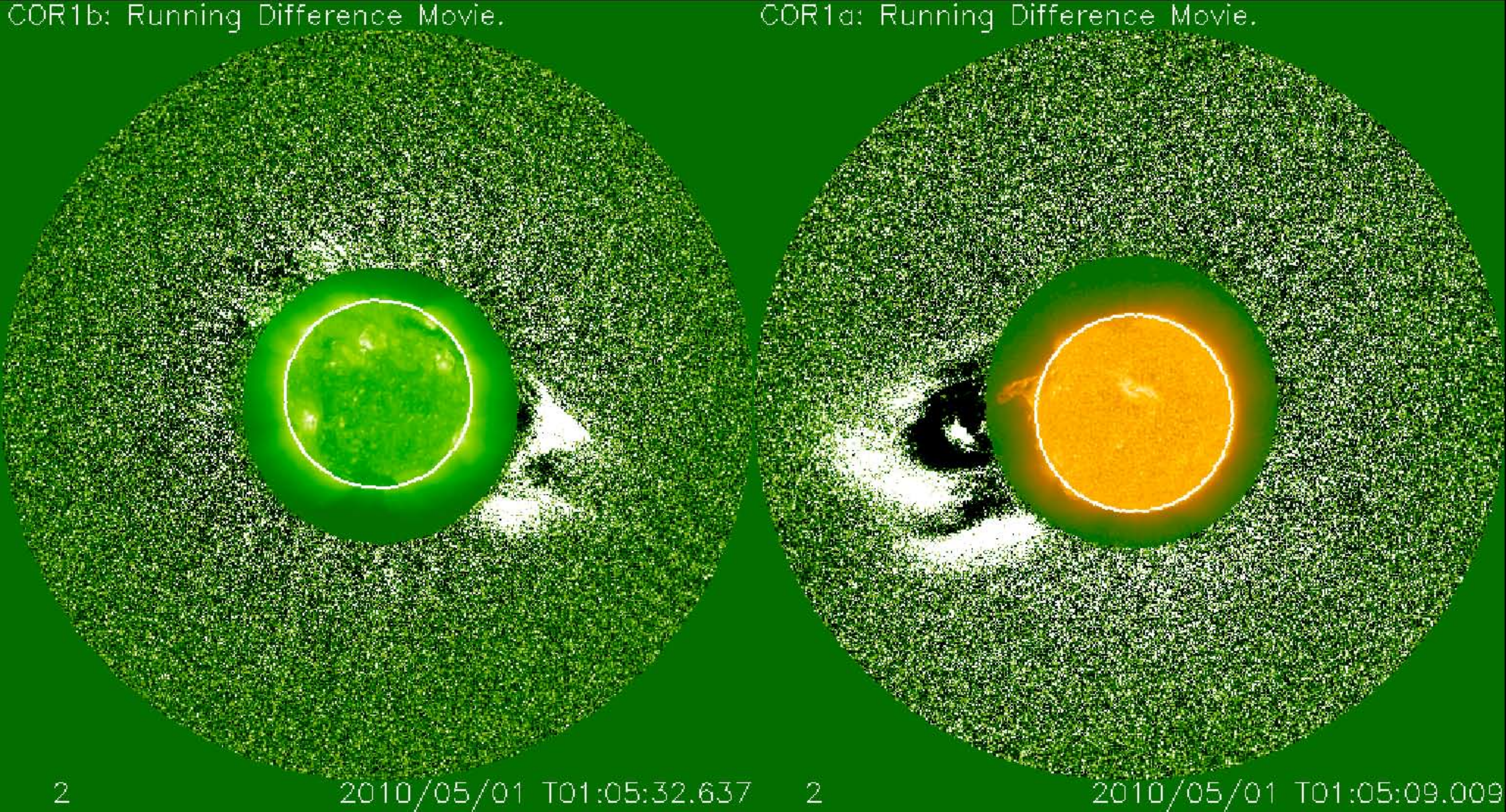}
\caption{ Difference  images of the CME observed on 1 May 2010. Left: STEREO-B/COR1 superimposed with EUVI 195 \AA\ ; Right:  Superposition of the STEREO-A/EUVI 304 \AA\  prominence at 00:06 UT and the STEREO-A/COR1 image of the CME at 01:05 UT. }
\label{20100501_COR1}
\end{figure}

\section{Radial eruptions with the roll effect}
Here we consider cases where there is still significant rolling of the prominence spine before the CME is formed, 
but subsequently the filament and CME propagate outwards radially.
Such cases are usually observed when the pre-eruptive filament is lying at the base of a pseudostreamer magnetic configuration where the arcades below the pseudostreamer have a much greater lateral extension relative to the pseudostreamer stalk which is essentially radial (an example of a non-radial pseudostreamer is shown in Figure 3).  The asymmetry in the filament position relative to the overlying coronal arcade might lead to additional force on the filament that could enhance the rolling motion of the erupting filament during the early stage of the eruption. This lateral motion of the filament is directed towards the null point in the magnetic configuration. The radial geometry of the pseudostreamer branches (defined as bundles of open magnetic field lines on opposite sides of the spine-fan projection on the sky) in this case favors the radial propagation for the CME flux rope formed in the later stages of the filament eruption. 

\subsection{1 August 2010 twin filaments sympathetic eruptions.}
\begin{figure}
\center
\includegraphics[scale=0.27]{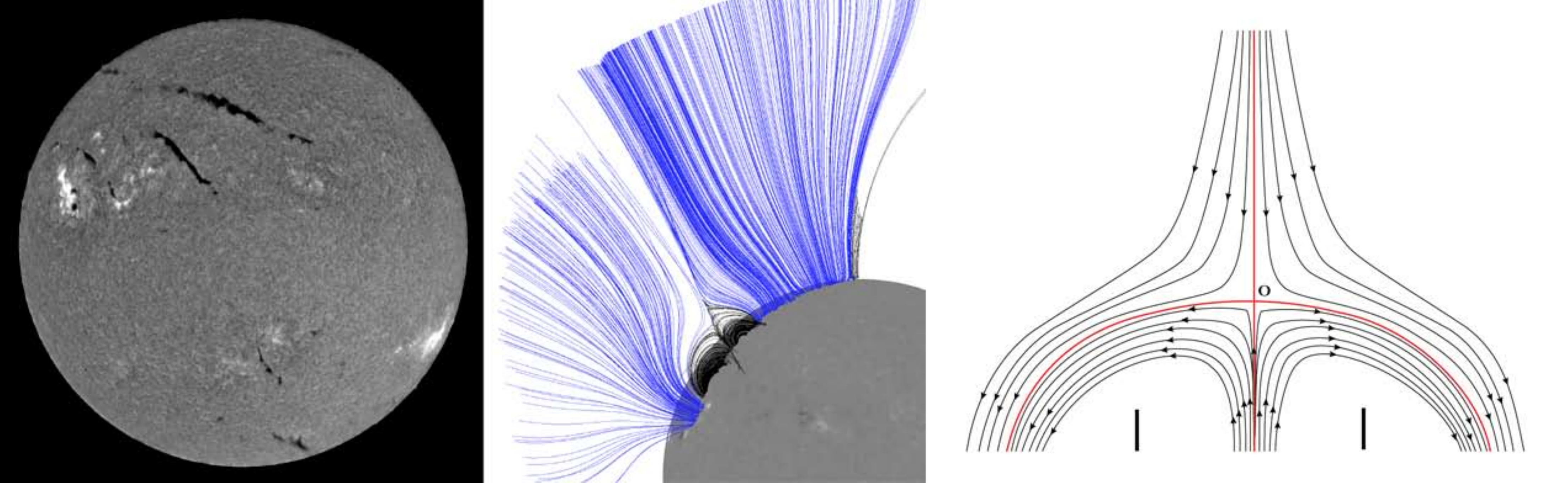}
\caption{ Left panel: BBSO H$\alpha$ image of the long \emph{twin filaments} in the northern hemisphere before they erupted on 1 August 2010. The eruptions were widely observed by SOHO, STEREO, SDO and ground-based observatories. The separation in time between the eruptions of the two filaments was 11h 30Õ. Both twin filaments (filaments with the same chirality) were lying at the base of a pseudostreamer; Middle panel: PFSS extrapolation of the pseudostreamer magnetic field above two dextral filament channels beneath the depicted coronal arcades shown in black on 30 July 2010: rotated to limb view;  Right panel: Magnetic configuration of the pseudostreamer and the position of the twin filaments (short vertical lines) relative to the null point, O, and magnetic separators (red lines).}
\label{2010Aug01}
\end{figure}

Pseudostreamers appear in unipolar regions above multiple polarity reversal boundaries. Some of these polarity reversal boundaries can be filament channels, and when this is the case they can occur as twin filament channels often containing twin filaments. 
Figure \ref{2010Aug01} shows twin filaments and the coronal magnetic field configuration above them on 2010 August 01. This event has been described in detail by Panasenco and Velli, 2010; Schrijver and Title, 2011; T{\"o}r{\"o}k et al., 2011. The slight rolling motion of the erupting filaments toward the null point in the pseudostreamer configuration has been observed and reported in T{\"o}r{\"o}k et al. (2011). Overall both CMEs, that originated under the pseudostreamer, propagated radially along the axis between the pseudostreamer branches.

\subsection{16 May 2007 south polar crown filament eruption.}

\begin{figure}
\center
\includegraphics[scale=0.2]{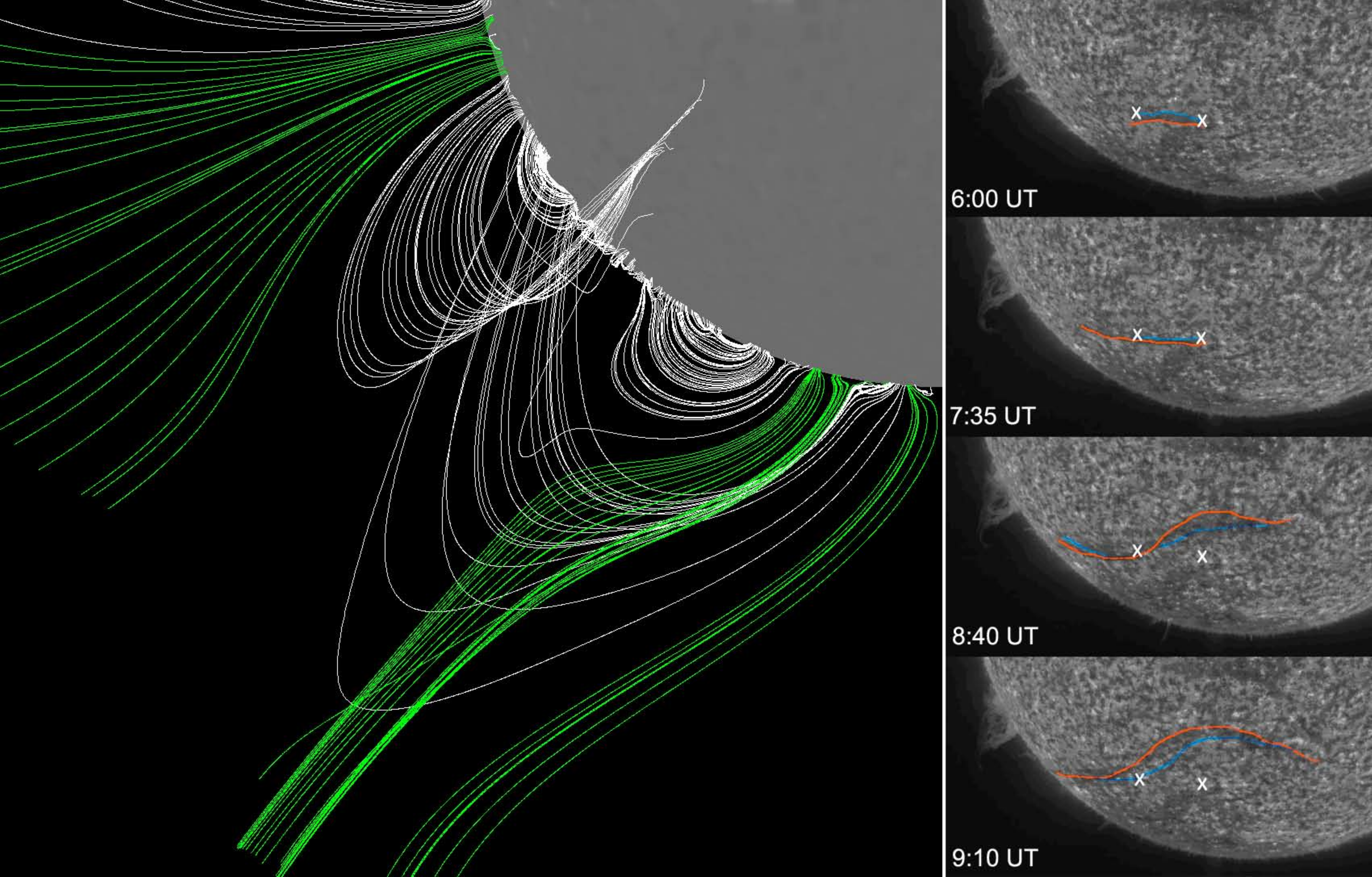}
\caption{ Left panel: PFSS extrapolation of the pseudostreamer magnetic field above the polar crown filament channel under the southern lobe of the pseudostreamer on 15 May 2007 06:06 UT rotated to limb view; Right panel: images from STEREO/SECCHI/EUVI showing the rolling motion of the erupting filament on 16 May 2007. From this perspective the highest part of the pre-eruptive filament is shown by the red line and the lowest border by the blue line. The rolling motion is revealed as the initially highest part of the filament gradually becomes the lowest part and the approximate lowest part in the  middle of the filament becomes the highest part. The roll appears to begin between the white Xs superposed on the blue line. It propagates in both directions away from the initial site (white Xs) and towards the apparent ends of the prominence as seen against the chromosphere.}
\label{2007May15}
\end{figure}

Figure \ref{2007May15} shows a PFSS extrapolation of the open field lines of the pseudostreamer and coronal loops overlying a filament under the southern  lobe of the pseudostreamer on 15 May 2007. On the right column are anaglyph images from EUVI showing an erupting filament on 16 May 2007;
the separation angle between STEREO-A and B is 8.14 degrees. To assist the reader, the upper edge of the filament is outlined by a red line and the lower edge by a light blue line. When one watches the 3D movies of this event created from STEREO/EUVI data, one can identify a rolling motion beginning to the east of the middle of the filament. A consequence of this rolling motion is that the bottom edge
of the filament crosses gradually over the top edge. The roll appears to begin between the white Xs superposed on the blue line. It propagates in both directions away from the initial site (white Xs) and towards the apparent ends of the prominence as seen against the chromosphere. We also see that this erupting filament is erupting non-radially and rolling northward in the direction of the pseudostreamer null point (left panel in Figure 13).

\section{Radial eruptions without the roll effect}

The final examples are cases where there is no visible rolling of the filament, and both the filament and CME propagate radially outwards through the corona.
 
\subsection{3D Reconstruction of the erupting prominence on 28 February 2010}

\begin{figure}
\center
\includegraphics[scale=0.2]{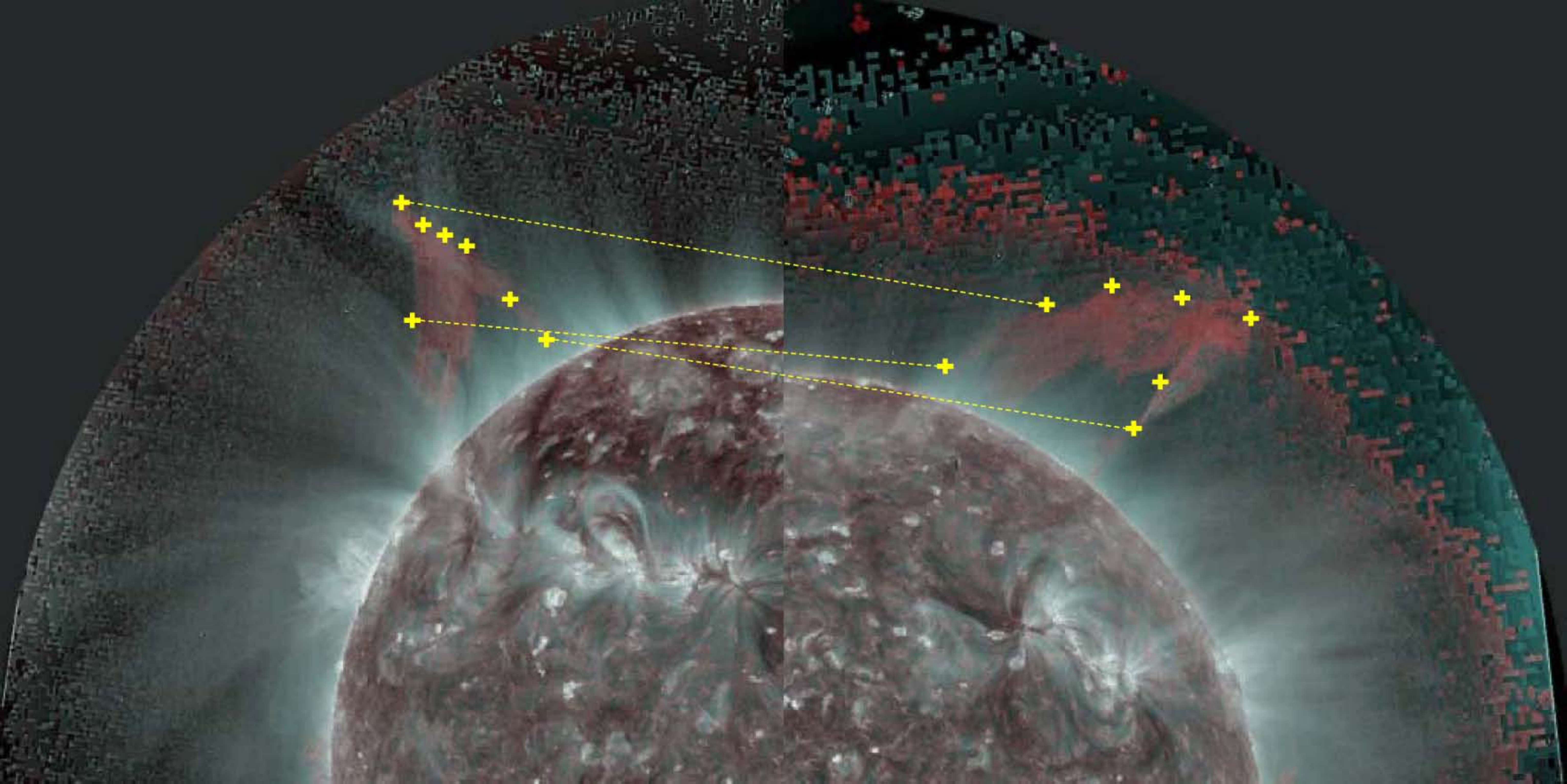}
\caption{ The 28 February 2010 eruption as seen by EUVI on STEREO-A (right) and -B (left). The images are composites of 304 \AA\ (red) and 195 \AA\ (silver) images. The time is 13:16 UT. The yellow stars mark the tie-pointing locations. The three yellow lines are visual aids to visualize the 3D configurations of the prominence. }
\label{Ang1}
\end{figure}

A polar crown prominence around the northern coronal hole (CH) erupted over the course of a day on 28 February 2010. The eruption was located on the far-side of the solar disk but it was captured by the EUVI imagers onboard both STEREO satellites. The two EUV viewpoints revealed a very different behavior of the erupting prominence during its ascent which is the reason we chose to analyse its 3D morphology in detail. In particular, EUVI-A shows a long piece of the polar crown prominence lifting off parallel to the surface. In contrast, the EUVI-B images show a very strongly kinked prominence (Figure \ref{Ang1}).

\begin{figure}
\center
\includegraphics[scale=0.2]{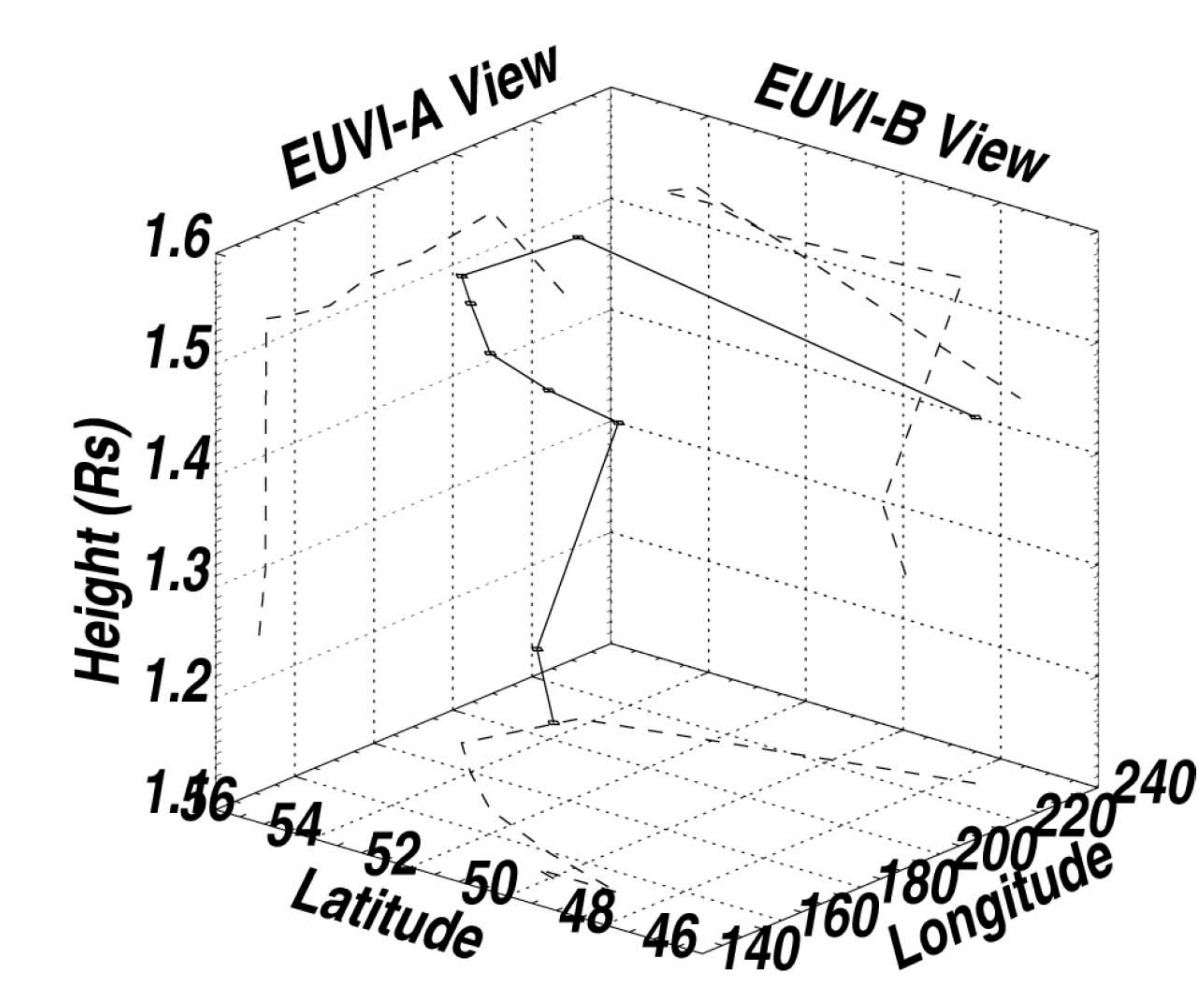}
\caption{ The 3D configuration of the prominence in the 28 February 2010 eruption using the points in Figure 14. The prominence has a small bend (about $6^\circ$ wide) that creates the appearance of a kinked structure in the EUVI-A view. Note that the footpoints of the structure are not visible and hence have not been measured.}
\label{Ang2}
\end{figure}

It appears, therefore, that the apparent kinked morphology from one viewpoint could be just a projection effect. Since we have observations from two viewpoints, we can quantify this effect because we can localize the prominence in 3D space. Tiepointing is the standard technique for this purpose and has been described before (Liewer et al., 2009; Liewer et al., 2011; Thompson, Kliem, and T{\"o}r{\"o}k, 2012). The technique relies of the correct identification of the same structure in the two images. Filaments are ideally suited for this because of their highly detailed morphology. Because we are interested in the 3D shape of the large scale structure of the prominence, we concentrate only on its 'backbone' in the EUVI 304 \AA\ images. We use relatively few points to determine its orientation. We select six points for tiepointing which are shown by the yellow crosses in Figure \ref{Ang1}. To aid the reader, we connect some of those points with the dashed lines. We use the IDL routine \textit{sunloop\/} which is part of the SECCHI software distribution in \textit{Solarsoft}. The routine allows the user to select one feature in one viewpoint and then click on the same point (according to the use) on the other image. The program returns the location of the feature in 3D space defined by its longitude, latitude and radial distance from the solar surface. Our results are shown in Figure \ref{Ang2}  along with the calculated projections of the prominence backbone for the EUVI-A, and B views and on the surface. 

\begin{landscape}
\begin{figure}
\center
\includegraphics[scale=0.3]{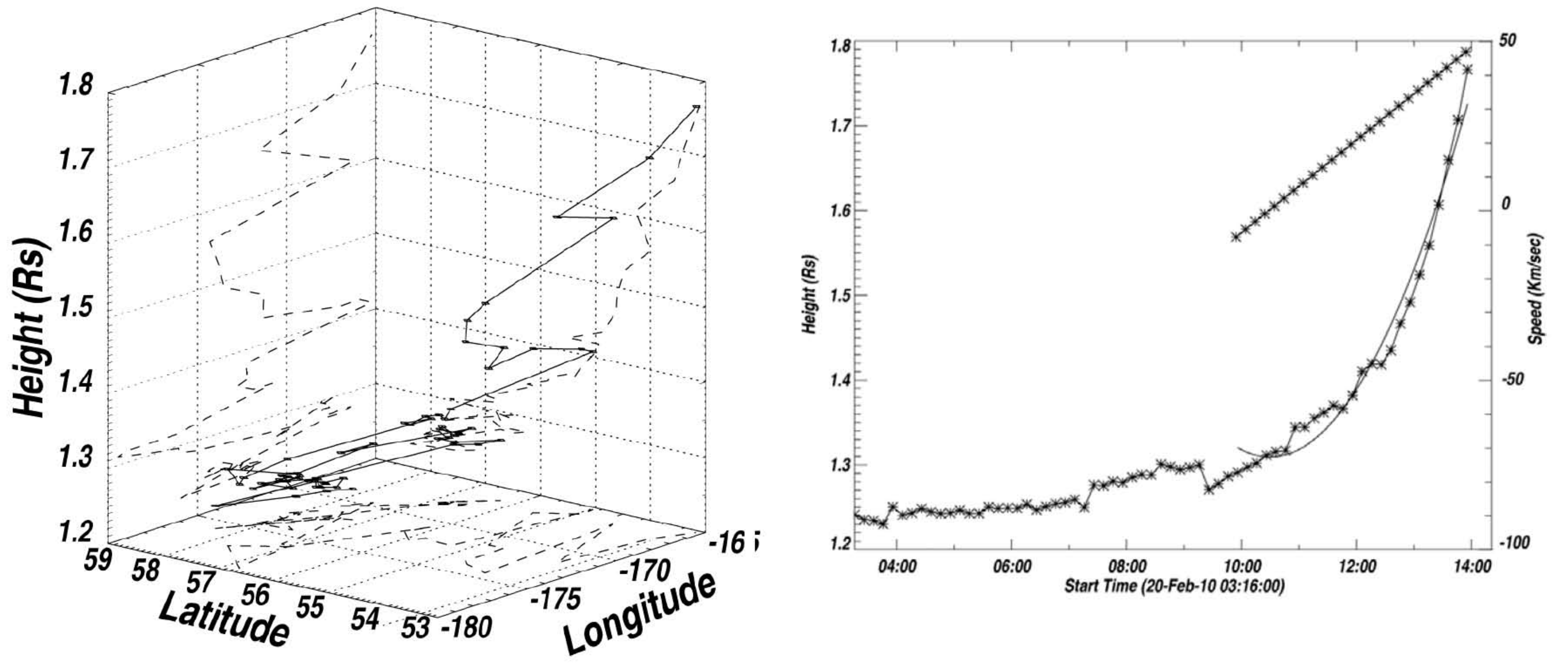}
\caption{ True radial height-time plots for the 28 February 2010 prominence eruption. The highest point of the structure was used. Left panel: 3D height-time plot of the prominence front. The eruption is deflected from radial by only $6^\circ$ in latitude. The longitudinal variation is likely the result of errors in the identification of the same feature in both EUVI views. Right: Fit to the height-time plot. Only points after 10:00 UT were used. The velocity curve is shown by the straight line with stars and the values are shown on the right-hand side of the plot.}
\label{Ang3}
\end{figure}
\end{landscape}

\begin{figure}
\center
\includegraphics[scale=0.2]{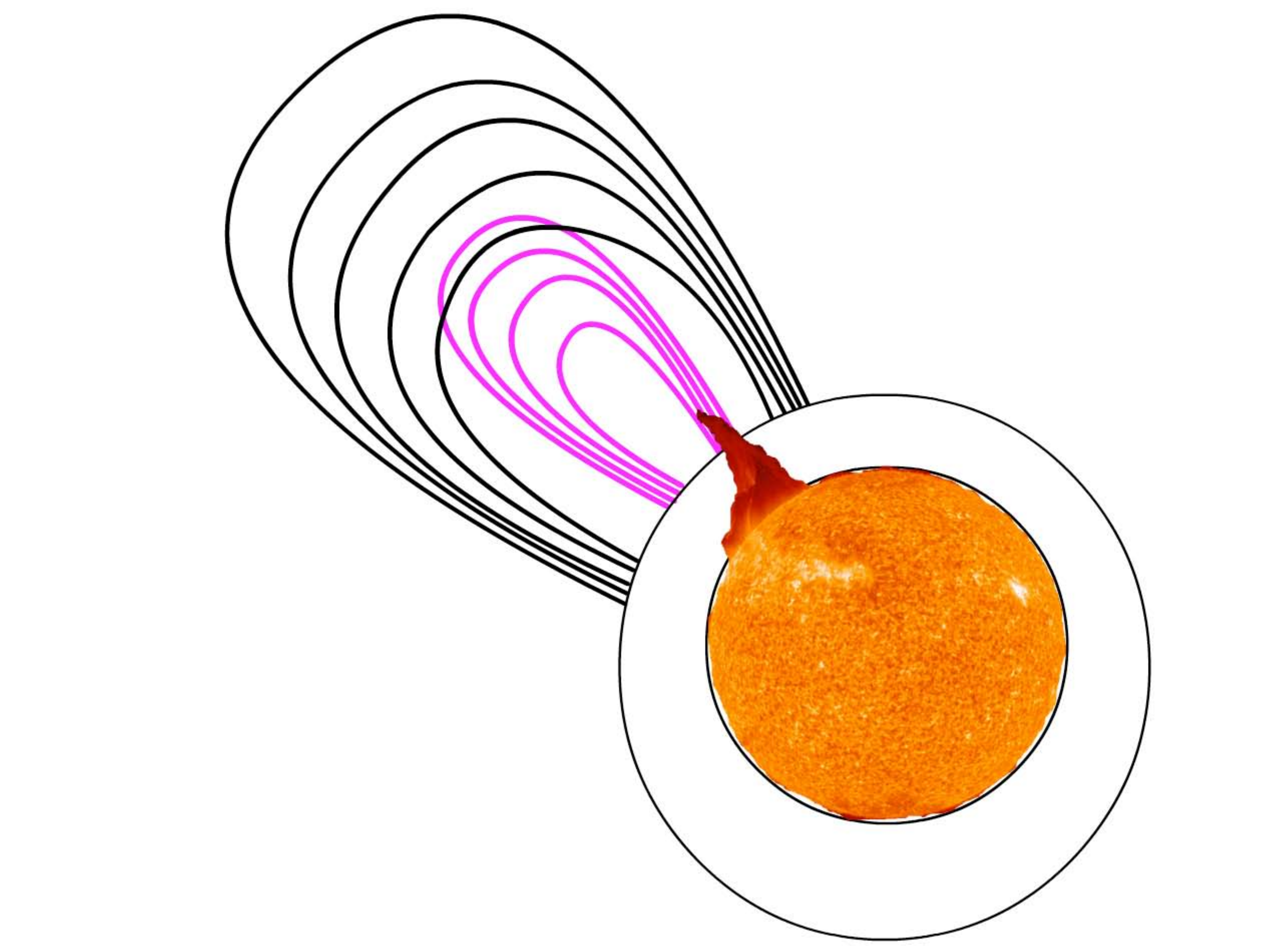}
\caption{ Radial eruption on 28 February 2010.  Superposition of the outer boundaries of CME (black lines)  and erupting prominence inside (magenta lines) as observed by STEREO-B/COR1 during different moments of the eruption and  the solar disk  with the prominence before eruption (STEREO-B/EUVI  304 \AA\ ) }
\label{2010Feb28}
\end{figure}

The measurements verify that the appearance of  a kink crossing is simply a projection effect. It is probably caused by the narrow shape of the prominence. The prominence is quite thin with a 70 deg longitudinal extent but only 8 deg of latitudinal extent. We cannot make a definitive measurement of the size of the prominence because its footpoints in the lower atmosphere are not visible from both EUVI telescopes. But we can easily trace the highest point along the backbone during the eruption. The resulting 3D height-time plots is shown in Figure \ref{Ang3} (left) where we can see that the prominence is erupting quite radially; the latitude changes by only 4 deg while the point remains within a 10 deg longitudinal range. The height-time profile  (Figure \ref{Ang3}, (right)) is quite typical for such slow eruptions. We measure a very slow rising during the first seven hours followed by a rapid increase after 10~UT. We fit the height-time plot only for the period of the fast rise using a 2nd degree polynomial. The resulting velocity profile is marked by the straight line of star symbols. We find that the prominence attains a speed of only 50 km s$^{-1}$ at 14~UT when it leaves the EUVI field of view. The prominence and the corresponding CME propagate radially outwards through the corona (Figure 17). 

In summary, the three-dimensional analysis of the 28 February 2010 prominence results in two important findings for this paper. First, we find that the kink morphology can be a result of projection effect which has important ramifications for past identification of kinked prominences. Second, we determine that the eruption is radial without any hint of equatorial deflection even though it is a slow prominent eruption (Figure 17). This is contrary to the prevailing consensus in the field which holds that prominence eruptions, especially during solar minimum, tend to be deflected towards the magnetic equator. Although this is certainly true for some events, it should not be taken for granted for any event. 

\subsection{Radial eruption on 6 July 2010.}

\begin{figure}
\center
\includegraphics[scale=0.18]{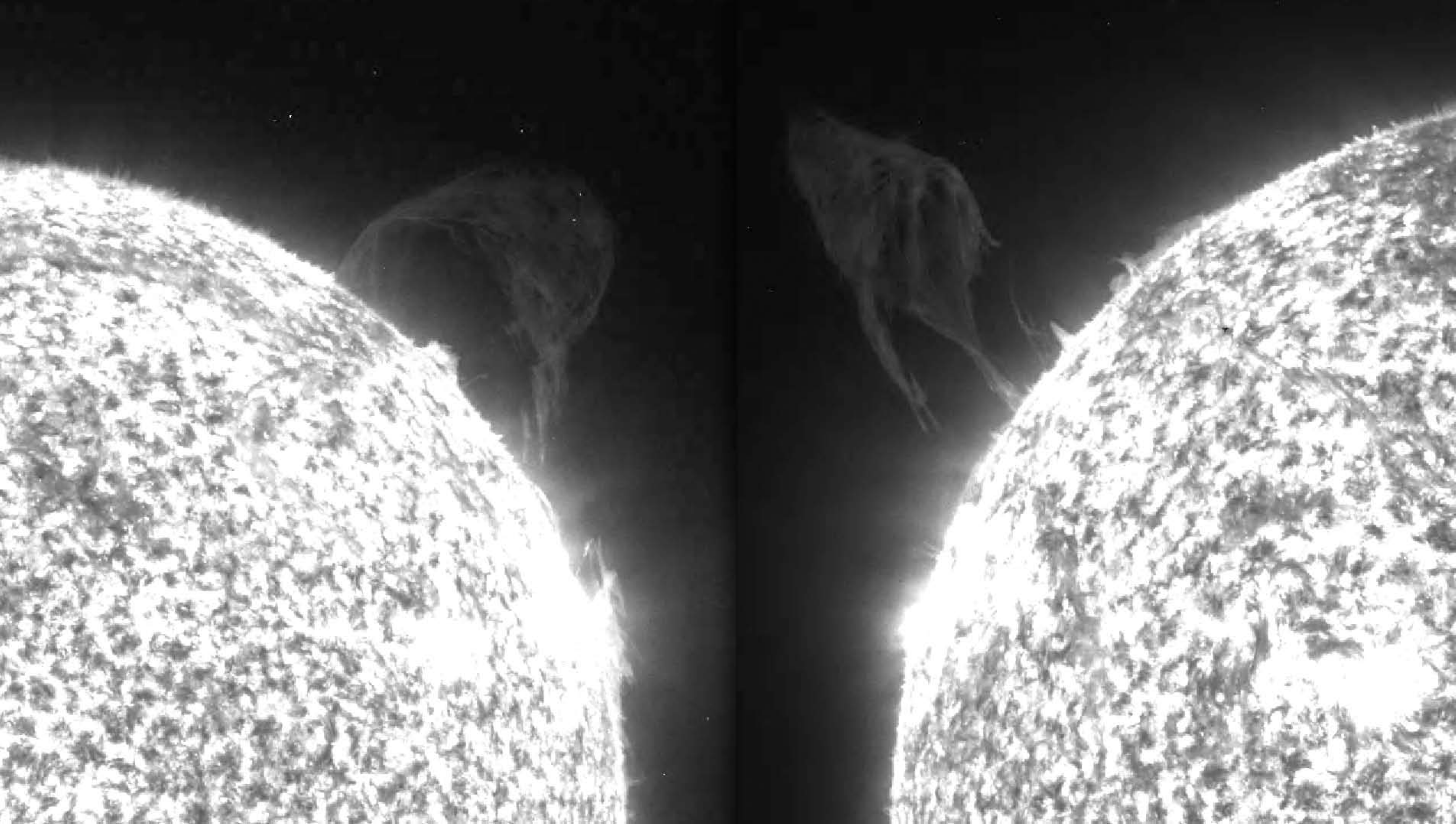}
\caption{ Radial eruption on 6 July 2010 seen from two perspectives.  Left and right figures are STEREO-B and STEREO-A/EUVI 304 \AA\ images captured during the eruption at 06:06:15 UT. }
\label{2010Jul06}
\end{figure}

A polar crown prominence around the northern CH erupted on 6 July 2010. The eruption was captured by the EUVI imagers onboard both STEREO satellites and SDO. The erupting prominence appeared to rise very radially and showed very little bending of the prominence spine during this eruption (Figure  \ref{2010Jul06}). In Figure 19  we also examined the position of the filament before its eruption on 6 July (SDO/AIA) beneath the coronal fields generated by the PFSS model. The filament is centered symmetrically beneath the arcade of overlying coronal loops. Therefore, as the filament rises, the magnetic pressure of the magnetized filament cavity and overlying arcade should be approximately equal on the two sides of the filament. The absence of the lateral rolling motion of the filament spine is consistent with the deduced presence of the magnetic force balance between the fields on the two sides of the prominence.
\begin{figure}
\center
\includegraphics[scale=.22]{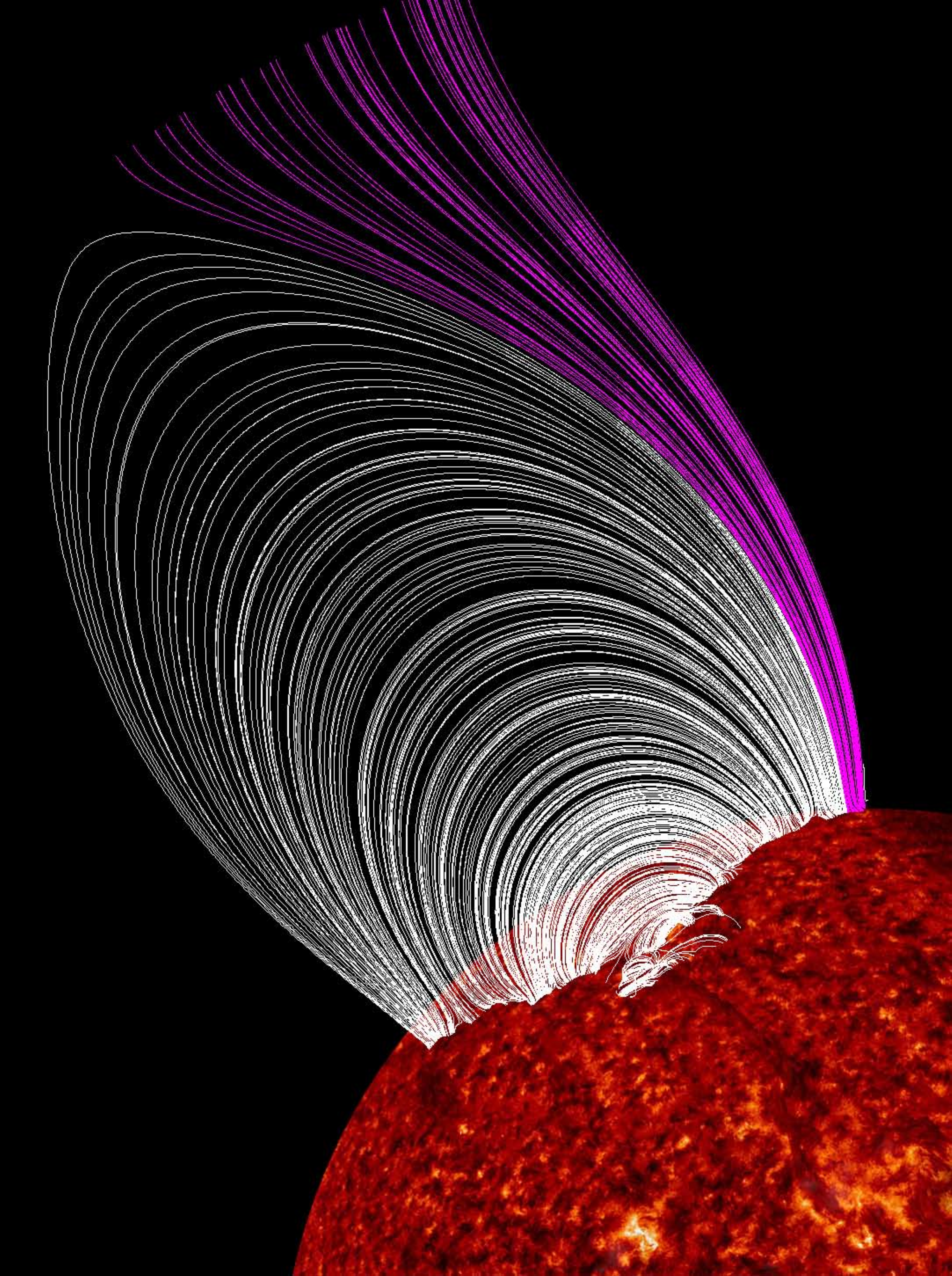}
\caption{  PFSS extrapolation of the open solar magnetic field and the coronal loops overlying the filament as observed by SDO 304 \AA\  on 6 July 2010 at 01:14 UT. The filament is centered symmetrically beneath the arcade of overlying coronal loops. }
\label{2010Jul06PFSS}
\end{figure}

\section{Discussion}

The path followed by the CME, and by the erupting filament immersed within it, depends on the macroscopic forces to which the erupting plasma is subject.
Both the roll effect and the variable CME trajectories imply that the energy in the background corona is non-negligible compared to that liberated in the ejection. In the opposite case of negligible background energy,  one would expect
an essentially radial expulsion with a self-similar expanding spherical shock wave \`a la Sedov (Sedov, 1981).  Describing the propagation of a part of a plasma cloud involves considerable difficulties, from the proper definition of the plasma cloud volume, to the separation of plasma cloud from the background field, to understanding the photospheric inertial effects of the parts of the plasma volume which remains connected to the Sun. Indeed, some or all of these problems are also intrinsic to numerical simulations. Consider then a volume of plasma as it is ejected and propagates outward into the solar corona. Parker (1957) focused on the possible shapes the plasma should take depending on the relative importance of kinetic and magnetic pressures, and then included the effects of the outward motion of the center of mass in a field with gradients. Assuming a general shape in the form of a deformable prolate spheroid, and using the virial theorem in its tensor form, he showed that if the internal pressure was dominated by the gas, rather than the magnetic field, the plasmoid would elongate into an infinitely thin shape aligned with the external magnetic field.  Pneuman and Cargill (1985) showed that if the plasma $\beta$ (ratio of kinetic to magnetic pressures) inside the body was very small, and it was initially in equilibrium with its surroundings, then it would move outward self-similarly, maintaining its shape. If we assume $\beta << 1$, the dominant force in determining the motion of the plasma center of mass will be the one arising from the magnetic field stresses, which, in the simplest case of a magnetically isolated plasma body, may be written as
\begin{equation}
{\bf F} = - \int _S {\bf dS} {B^2_e\over 8 \pi},
\end{equation}
where $S$ is the surface bounding the volume of the ejected material and $B_e$ is the external magnetic field, as perturbed by the presence of the ejecting plasma itself. More generally the global force
has a contribution coming from the non-diagonal part of the magnetic stress (magnetic tension) whose role is to eliminate the net force parallel to the magnetic field connecting the ejecta to the
(arbitrarily defined at this point) outer magnetic field,
\begin{equation}
{\bf F}_i = - \int _S d{\bf S} {B^2_e\over 8 \pi} + \int _S {{\bf B} {\bf B}\cdot d {\bf S} \over 4 \pi},
\end{equation}
Parker (1957), Pneuman and Cargill (1985). For the cases of simple radial magnetic field background, these expressions lead to the so-called melon seed expulsion of diamagnetic plasmoids (see also Rappazzo et al. 2005 for an application to blobs formed at the tips of helmet streamers).

\begin{figure}
\center
\includegraphics[scale=0.3]{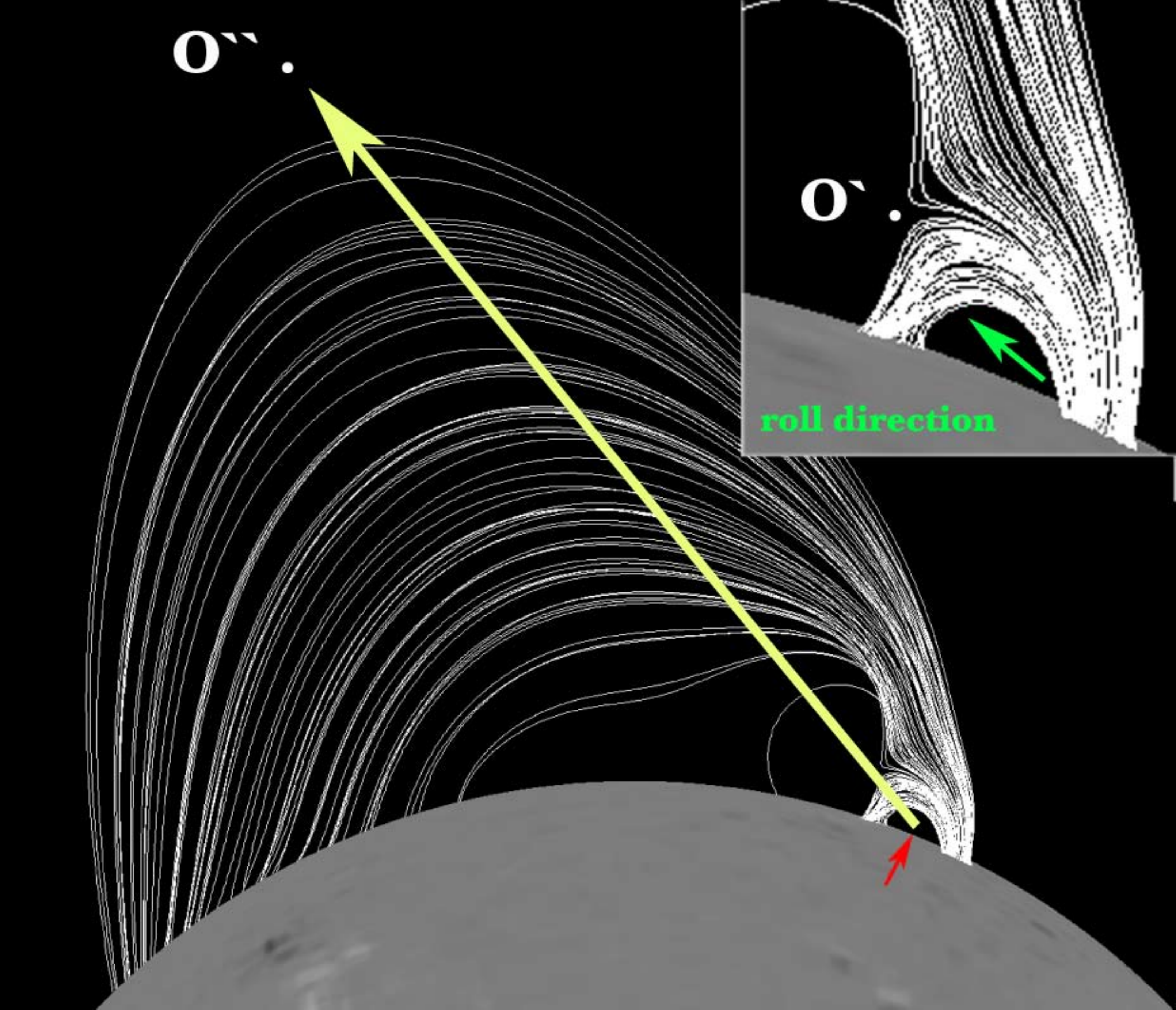}
\caption{The PFSS extrapolation of the coronal loops overlying the filament channel indicated by the red arrow (30 April 2010; see Figures 10-11 for details). The green arrow indicates the direction of the rolling motion of the erupting filament,  the yellow arrow indicates the direction of the CME deflection. O` and O`` are local and global null points.}
\label{discussion}
\end{figure}

These expressions are important because they show that regions in the corona separating strong and weak magnetic field regions, for example the boundaries of coronal holes, will act as walls repelling the motion of plasmoids from the strong field regions, whereas the neighborhoods of coronal null points (for example, between spine and fan for a pseudostreamer configuration), will be regions {\it towards} which an ejecting plasma will be accelerated, as they are regions of vanishingly small magnetic pressure. The dynamics of course is richer, and nonlinearly so, because as a consequence of such motions regions neighboring null points may develop strong current sheets, leading to attraction along a privileged axis (locally orthogonal to the tearing sheet)   and repulsive in the plane of the sheet itself (a representative example is given by the sympathetic eruptions simulated in T{\"o}r{\"o}k et al. 2011, where the breakout of one filament at first inhibits that of the other,
but once sufficient flux is removed, the breakout of the second filament also follows). A detailed analysis of the acceleration of the filament and global CME requires separating the components of an eruption. Consider then the global CME as a large-scale plasmoid bubble, dominated by the cavity magnetic field, elongated along the filament axis. In this case one may apply the simplifications of a magnetically isolated system, compared to the rest of the coronal field, which in the case of
a CME with small dimensions compared to the global coronal scales would lead to an expression for the
force as something like
\begin{equation}
{\bf F} = -c V \nabla {B^2\over 8 \pi}
\end{equation}
with $V$ the volume of the CME and $c$ a constant $1.5 > c > 1$.
This can explain departures from radial expansion due, for example, to the presence of a coronal hole, which occurs on large scales. For the filament itself, local deflections occuring in the neighborhood of the relatively low-lying null points in pseudo-streamer or more general multipolar configurations, and complexities due to field lines not isolating the plasma entirely need to be taken into account. Differences in the acceleration of the filament (or filament spine) and the overall CME could therefore  arise quite naturally in the initial phases of the eruption. More generally the expansion of the filament leads not only to macroscopic forces but also moments tending to rotate the prominence spine (responsible for the roll effect). Here the natural response of the filament should lead to deflection and alignment following the neighboring cavity field. As shown in the PFSS extrapolations, the asymmetries in the overlying arcade field correspond to both roll and deflection (Figure \ref{discussion}). 

When multiple arcades are present below a pseudo-streamer or a streamer configuration, the various polarity regions form a hierarchy of null points (and possible lines/fan structures). From the PFSS modeling we found that the lateral rolling motion of the filament is directed first towards the closest, \emph{local} null point above the (asymmetric) coronal arcade before this arcade reconnects creating the CME flux rope. Once reconnection has occurred, the just-born CME flux rope then follows moving non-radially towards the higher, more distant \emph{global} null points/lines separating larger scale structures (the fan/spine null point of the large-scale pseudo-streamer, or the tip of the helmet streamer for a globally bipolar configuration).

\section{Conclusions}

We have considered the global environment around CMEs, as can be approximated by the Potential Field Source Surface (PFSS) model. From our analyses, we have demonstrated the following relationships for eruptive solar events consisting of an erupting filament surrounded by a cavity and an enveloping CME:

\begin{itemize}
\item Both erupting filaments and their surrounding CMEs are non-radial only in the direction away from a nearby coronal hole and toward local and global null points. Due to the presence of the coronal hole, the global forces on the CME are asymmetric. The CME propagates non-radially in the direction of least resistance  as we demonstrate by comparing low latitude and high latitude examples.
\item The presence of lateral rolling motion in erupting filaments can account for their observed bending and twisting during eruption. 
\item The roll effect can depend, at least in part, on the asymmetries in the magnetic flux densities on the two sides of a polarity reversal boundary in a filament channel under a rising coronal loop system. 
\item The magnitude of the filament deflection from the radial propagation during its eruption is typically greater than the deflection of the corresponding CME. This difference can be explained by the different origins and character of filaments and CMEs: (a) The spine of filaments exist as thin ribbons before their eruption and these ribbons roll, bend, and twist in characteristic ways during eruption.  (b) Before their eruption, CME magnetic fields are tunnel-like systems of many fine coronal loops, a topology that allows an ascending loop system to reconnect and form a flux rope via magnetic reconnection below the associated and ascending filament ribbon.  Such flux ropes therefore encapsulate the eruptive filaments, which still have some freedom to keep moving along the original direction of eruption, as defined prior to the flux rope formation.
\item We confirm previous findings that the CME phase of rapid acceleration is initiated concurrently with the magnetic reconnection of a rising coronal loop system.
\item In events with the roll effect and non-radial filament eruptions, but with radial CMEs, rolling motion and lateral deflection for the erupting filament is observed only during the early stages of the eruption. 
\item CME flux ropes commonly propagate radially for eruptive events associated with   pseudostreamers. 
\item Non-radial and radial filament eruptions often show kinked shapes, which may, at least in part, be a consequence of projection effects from the rising ribbons of the filaments together with large scale rolling motions. 
\end{itemize}
	
Through modeling and comparison with observed events, we anticipate that major
twists and non-radial motions in erupting prominences and CMEs will become predictable to the extent that their environments are well-defined and measurable.

\begin{acks}
We are indebted to the SOHO, STEREO/SECCHI  and  SDO teams. 
O. P. and S. M. are supported in this research under the NASA grant NNX09AG27G.  The work of M.V. was conducted at the Jet Propulsion Laboratory, California Institute of Technology under a contract from the National Aeronautics and Space Administration.  A.V. is supported by NASA contract S-136361-Y to the Naval Research Laboratory. SOHO is a mission of international cooperation between ESA and NASA. The SECCHI data are produced by an international consortium of the NRL, LMSAL and NASA GSFC (USA), RAL and Univ. Bham (UK), MPS (Germany), CSL (Belgium), IOTA and IAS (France).The AIA data used here are courtesy of SDO (NASA) and the AIA consortium.
\end{acks}

\end{article}
\end{document}